\documentclass[12pt]{article}
\usepackage[utf8]{inputenc}

\usepackage{setspace}

\usepackage[section]{placeins}

\usepackage[top=1.25in, bottom=1.25in, left=1.25in, right=1.25in]{geometry}

\newcommand{\nofootnote}[1]{%
  \begingroup\def\thefootnote{}\footnotetext{#1}\endgroup}

\usepackage[round]{natbib}
\newcommand{\possessivecite}[1]{\citeauthor{#1}'s (\citeyear{#1})}

\usepackage[usenames,dvipsnames]{xcolor}
\usepackage[colorlinks=true,allcolors=Black,bookmarks=true]{hyperref} % Add the option , backref if you want links back to the page of the citation.

\usepackage{amsmath}
\usepackage{amsthm}
\usepackage{amsfonts}
\usepackage{thmtools}
\usepackage{amssymb}

\theoremstyle{plain}
\newtheorem{theorem}{Theorem}
\newtheorem{proposition}{Proposition}
\newtheorem{claim}{Claim}
\newtheorem{lemma}{Lemma}

\theoremstyle{remark}

\theoremstyle{definition}

\declaretheorem[style=definition,qed=$\triangle$]{example}

\usepackage{graphicx}
\usepackage{caption}
\newcommand\fignote[1]{\captionsetup{font=small}\caption*{#1}}
\usepackage{subcaption} % For sub-figures instead of subfig
\usepackage{tikz}
\usetikzlibrary{positioning}
\usetikzlibrary{arrows}

\usepackage{tkz-graph}

\usepackage{pgfplots}
\pgfplotsset{width=10cm,compat=1.9}

\usepackage{booktabs}
\usepackage{tabulary}
\usepackage{lscape}
\usepackage{threeparttable} % Table notes

\usepackage{pdfpages} % To insert external pdf as pages

\usepackage{afterpage}

\title{Friend-Based Ranking%
\nofootnote{We thank Abhijit Banerjee, Leonie Baumann, Yann Bramoull\'{e}, Bhaskar Dutta, Piotr Dworczak, Matt Elliott, Jeff Ely, Ben Golub, Jonathan Guryan, Sanjeev Goyal, Matt Jackson, Matt Leister, Neil Lloyd, Vesall Nourani, Brian Rogers, Agnieszka Rusinowska, Jonathan Weinstein, Leeat Yaariv and Yves Z\'{e}nou for helpful discussions on the project. We are also grateful to seminar participants at several institutions for their comments. This work was partly supported by the Agence Nationale de la Recherche under grant ANR13BSHS10010.}%
}
\author{Francis Bloch\thanks{ Paris School of Economics, Universit\'{e} Paris 1 Panth\'{e}on-Sorbonne, francis.bloch@univ-paris1.fr}%
\and Matthew Olckers\thanks{ Monash University, matthew.olckers@monash.edu}}
\date{\today\\
{$\ \ $}\\
}

\begin{document}

\maketitle

%%%%%%%%%%%%%%%%%%%%%%%%%%%%%%%%%%%%%%%%%%%%%%%
%%%%%%%%%% Abstract and contents
%%%%%%%%%%%%%%%%%%%%%%%%%%%%%%%%%%%%%%%%%%%%%%%
\singlespacing

% Must be less than 100 words.
\begin{abstract}
We analyze the design of a mechanism to extract a ranking of individuals according to a unidimensional characteristic, such as ability or need. Individuals, connected on a social network, only have local information about the ranking. We show that a planner can construct an ex post incentive compatible and efficient mechanism if and only if every pair of friends has a friend in common. We characterize the windmill network as the sparsest social network for which the planner can always construct a complete ranking.
\end{abstract}

\bigskip


\medskip


% \newpage

% \tableofcontents

\pagebreak%

\onehalfspacing

%=============================================================
\section{Introduction}
%-------------------------------------------------------------

In many social networks, individuals gradually acquire information about their friends through repeated interactions. Pupils in a class learn about the ability of other pupils with whom they write joint projects, workers in a firm learn about the productivity of the coworkers in their teams, and members of a community learn about the needs of their neighbors. This information, which is scattered in the social network, may be of great use to an external planner who wants to extract information about members of the community. A teacher wants to learn about the ability of her pupils; an employer, the productivity of her workers; a funding agency, the most needy members of a community.

In the classical literature on mechanism design, the principal designs a mechanism which asks individuals to report on their own types. However, in some situations, it is impossible to have individuals truthfully report on their own types without relying on undesirable punishments. An alternative is to generate a mechanism which asks individuals  to report not on their own type but on the type of their friends in the social network. Pupils are asked to assess the performance of other pupils, workers are asked to measure the productivity of their coworkers, or villagers are asked to rank neighbors in the community. The objective of our paper is to analyze these mechanisms, that ask individuals to report about their friends in the social network---mechanisms that we term ``friend-based ranking mechanisms". In particular, we want to understand how the architecture of the fixed social network affects the planner's ability to construct a mechanism having desirable properties.

%--- this paragraph is new ----------------------
While we treat the social network as exogenous, we note that in some situations the planner can design the network.  The editors of scientific journals choose the set of referees for every paper. Referees are themselves authors and hence compete for space in the journals. Conference organizers and funding agencies also choose the assignment of projects to reviewers who have submitted papers or projects.\footnote{While funding agencies typically rely on outside experts, the National Science Foundation, in order  to reduce the burden on outside referees,  has experimented with new peer-review systems asking applicants to rank other applicants.} In large-scale online courses, as teaching staff time and resources are limited,  grading is distributed among the students. The design of peer-grading systems is an important issue for the success of online courses.

We study a setting with three characteristics. First, we assume that information is local. An individual may make comparisons only among his direct friends. Second, information is ordinal. Individuals lack the ability to quantify characteristics and can only assess whether one individual has a higher characteristic than another. Third, we assume that the planner has only one instrument at her disposal: she constructs a complete ranking of the members of the community. Hence, the number of outcomes that the planner can select from is very restricted. The planner cannot use transfers, and cannot punish individuals by excluding them from the ranking. In particular, she cannot impose punishments for inconsistent reports, as in the classical literature on implementation with correlated types \citep*{cremer1985optimal}.

We require that the planner's mechanism  satisfy two properties. First, individuals must have an incentive to report information truthfully. In the ordinal setting we consider, the natural choices for implementation concepts are dominant-strategy and ex post implementation. However, we notice that the limited number of outcomes in our setting means that dominant-strategy implementation is too strong, leading to impossibility results. We adopt instead ex post incentive compatibility as the desirable incentive property of the mechanism.

Second, we require the mechanism to be ex post efficient from the point of view of the planner, whose objective is to recover the true ranking of individuals in the community. More precisely, the ranking chosen by the planner must match the ranking that society would construct by aggregating all local information. If society can construct a complete ranking of individuals for any realization of types (a situation we label ``completely informative"), the ranking chosen by the planner must match the true ranking. Otherwise, when, for some type realizations, the structure of the network prevents a complete ranking, the  ranking chosen by the planner will be a completion of the partial order that the community is able to construct, and this completion will involve an arbitrary ranking across individuals who cannot be compared.

We first analyze mechanism design in completely informative societies. We show that a society is {\em completely informative} if every pair of individuals can be compared, either through ``self-comparisons" (the two individuals involved in the pair compare each other) or through ``friend-based comparisons" (a third individual observes both individuals  in the pair). Our main theorem shows that self-comparisons can be used only  if they are  backed by the comparison of a third individual. A mechanism satisfying ex post incentive compatibility and efficiency exists if and only if every pair of individuals has a common friend. Relying on reports by third parties may, at first glance, seem to be an obvious way to avoid inconsistent reports and elicit a truthful complete ranking. Note however that we also need to guarantee that the third party, the common friend, cannot manipulate his report to improve his own rank. This requires the planner to identify potential deviations from third parties, and results in a complex mechanism construction.

We then characterize the sparsest network which satisfies this property. When the number of  individuals is odd, this is the ``friendship network" of \citet*{erdos1966problem}, the only network in which every pair of individuals has only one common friend. This network, also known as the windmill, has one individual as a hub who connects all other individuals who form pairs. When the number of individuals is even, this is a variant of the windmill in which one of the ``sails'' contains three individuals instead of two. This network, while being sparse in the number of required comparisons, is very asymmetric and places a huge burden on one individual---the hub---who is responsible for a large number of comparisons.

We then turn our attention to societies which are not completely informative. We generalize our main theorem to show that the planner can construct an ex post incentive-compatible and efficient mechanism if and only if every pair of friends has a friend in common. Following the terminology of  \cite{jackson2012social}, every link must be ``supported" by a link to a third individual.\footnote{In \cite{jackson2012social}, this architecture is needed to support cooperation in a repeated model of favor exchange. The reason for which this architecture emerges in our setting is very different.} This requirement is different from the condition in completely informative societies, where all pairs of individuals (not necessarily linked) must be observed by a third individual.

If the network does not satisfy the condition for the construction of an ex post incentive-compatible and efficient mechanism, we can still hope to elicit comparisons among a smaller subset of individuals. To this end, we need to discard self-comparisons.  When the planner only relies on friend-based comparisons we construct a \emph{comparison network} by linking two individuals if and only if they have a common friend. We find that there exist two network architectures for which the planner can construct a mechanism satisfying  ex post incentive compatibility and efficiency. In the first architecture, the social network is bipartite (which is equivalent to the comparison network being disconnected).\footnote{A connected social network is \textit{bipartite} if the individuals can be divided into two disjoint sets $A$ and $B$ such that individuals in $A$ only link to individuals in $B$ and individuals in $B$ only link to individuals in $A$.} We use the bipartite structure to partition the set of individuals, so that individuals in one group rank individuals in the other group, and individuals  are ranked across groups in an arbitrary way. For these bipartite networks, truth telling is not only ex post incentive compatible, but also strategy-proof.  In the second architecture, all links form triangles, and we can use the general characterization theorem to prove existence of a mechanism satisfying ex post incentive compatibility and efficiency. However, we also note that even when we disregard self-comparisons, there exist network architectures for which mechanisms satisfying ex post incentive compatibility and efficiency cannot be constructed. The simplest example is a network of four individuals with one triangle and one additional link.

We then check if the requirements of our theoretical results are met by real social networks. We use social network data from villages in Karnataka, India, and hamlets in Indonesia. We check both the stronger requirement that every pair of \emph{individuals} has a friend in common and the weaker requirement that every pair of \emph{friends} has a friend in common. We find that the requirements are more likely to be met in the Indonesian networks than in the Indian networks as Indonesian networks are smaller than Indian networks. Out of the 633 Indonesian networks, 45 satisfy the stronger requirement and 127 satisfy the weaker requirement. None of the Indian networks satisfy the requirements. When the requirements are not satisfied exactly, comparisons may still be extracted by using the largest subnetwork which satisfies the requirements. We compare the share of comparisons retained by a partition mechanism, which ignores within group links in a partition of the network into two groups, to the share of comparisons retained by a support mechanism, which ignores unsupported links. The support mechanism retains strictly more comparisons in 162 of the 213 Indonesian networks for which we complete the computation.

Finally, we consider different variants and robustness checks of our model. We show that dominant-strategy implementation is too strong, leading to an impossibility result in triangles. We analyze the robustness of our mechanism to joint deviations by groups. We study whether coarser rankings are easier or harder to implement than complete rankings. We study the impact of homophily. If individuals of similar characteristics are more likely to form friendships, the planner is more likely to extract the necessary and sufficient friend-based comparisons to find the complete ranking.

We believe that our analysis can help a planner design a concrete mechanism in real world situations when side-payments are not possible.\footnote{When side-payments are possible, other mechanisms are available, like the ``Bayesian truth serum" of \citet*{prelec2004bayesian} or the peer-prediction mechanism of \citet*{miller2005eliciting}.} If the social network is exogenous, as in community targeting in villages or information elicitation among co-workers, our analysis identifies the network architectures for which desirable mechanisms can or cannot be constructed. Our analysis emphasizes the importance of triangles, and the need to ask third parties to check on self-reports. Our analysis also provides a careful construction of a mechanism which prevents third parties from manipulating their reports to improve their rank.

Perhaps more importantly, our analysis can help a planner design endogenous networks to assign projects to reviewers in peer-review and peer-grading systems. If the objective is to implement a dominant strategy mechanism, our analysis suggests to partition the set of individuals into different groups and ask individuals in one group to rank those in another. This will result in bipartite networks, leading to  a truthful and efficient ranking, but the price to be paid is that inter-rankings among individuals in the two groups will be arbitrary. If one looks instead for ex post incentive compatible mechanisms, our analysis indicates that peer-review  must involve the formation of cycles of order three (triangles). Asking pairs of individuals to evaluate each other, or constructing cycles of graders of length greater than three results in incentive problems and should be avoided. The total number of triangles formed, and the arrangement of triangles in the social network involves a trade-off between the burden assigned to every single reviewer and the total number of reviews. In order to economize on resources, one may want to construct a windmill network and  delegate a large number of comparisons to a single reviewer. In order to distribute the burden among reviewers equally, one may want to increase the number of triangles and construct a more balanced network.

\subsection{Literature Review}

We first discuss the relationship of our paper with the literature on community-based targeting in development economics. In community-based targeting, members of the community gather at a meeting to decide on a ranking of need to determine who will receive a social grant. The public nature of the meeting encourages truthful reports, but only if the number of grants is limited \citep*{rai2002targeting}. \citet*{alatas2012targeting} report on a field experiment in Indonesia in which community-based targeting was compared to proxy-means testing, another popular method of targeting. Community-based targeting captures a broader concept of welfare than can be captured by consumption measures and results in greater community satisfaction with the ranking. \citet*{alatas2016network} use network data collected during the same experiment to show that the information people have about fellow community members decreases sharply with social distance. As a result, networks with high density have less targeting errors from community-based targeting (relative to a proxy-means test) than networks with low density.

In contrast to community-based targeting, friend-based ranking mechanisms do not require the community to gather at a central meeting to aggregate their knowledge. In friend-based ranking, the planner asks each individual to rank his friends and aggregates the information herself. In a recent field experiment in Maharashtra, India, \citet*{hussam2017targeting} employ a friend-based ranking mechanism to rank entrepreneurs according to marginal returns to capital. The researchers divided entrepreneurs into groups of four to six according to geographic proximity and asked each entrepreneur to rank his or her fellow entrepreneurs. Despite entrepreneurs being placed in small non-overlapping groups, the ranking exercise was more effective at predicting marginal returns to capital than a machine-learning algorithm using a wide variety of survey measures.

Our analysis highlights the importance of a common friend to construct a truthful ranking of community members. In a different setting, the existence of common friends has also been noted as a way to increase informal transfers inside communities.  \citet*{karlan2009trust} show that common friends increase the borrowing possibilities in a setting where transfers are guaranteed by social collateral. \citet*{jackson2012social} demonstrate that networks where every link is supported by a link to a common friend are conducive to the exchange of favors.

The theoretical analysis of the paper is closely related to the literature in computer science and social choice theory studying peer selection. \citet*{alon2011sum} analyze the design of mechanisms to select a group of $k$ individuals among their peers.   \citet*{alon2011sum} prove a strong negative result: no deterministic efficient strategy-proof mechanism exists. Approximately efficient, stochastic, impartial mechanisms can be constructed, which are based on the random partition of individuals into clusters of fixed size such that individuals inside a cluster rank individuals outside the cluster.  \citet*{holzman2013impartial} analyze impartial voting rules when individuals nominate a single individual for office.  They identify a class of desirable voting rules as two-step  mechanisms, by which voters are first partitioned into districts which elect local winners, who then compete against one another to select the global winner. However, \citet*{holzman2013impartial} also highlight a number of impossibility results, showing that there is no impartial voting procedure which treats voters symmetrically, nor any impartial voting procedure which guarantees (i) that an individual whom nobody considers best will never be elected and (ii) that an individual whom everybody considers best will always be elected. \citet*{kurokawa2015impartial} and \citet*{aziz2016strategyproof,aziz2019} improve on the partition algorithm proposed in \citet*{alon2011sum}. They consider a more general setting, inspired by the pilot experiment  by the National Science Foundation to fund the Sensors and Sensing System program in 2013, following a suggestion by \citet*{merrifield2009telescope}. \citet*{kurokawa2015impartial} propose the ``credible subset mechanism'',  a process which first identifies candidates who are likely to win, and assigns ratings only to these candidates. \citet*{aziz2016strategyproof} propose a mechanism combining the insights of the partition mechanism of \citet*{alon2011sum} with the impartial ``divide the dollar" mechanism of \citet*{de2008impartial}. In a recent paper, \citet*{kahng2018ranking} extend the model of peer selection to construct a complete ranking of the individuals as we do. They analyze three different randomized algorithms, one where individuals are partitioned into $k$ groups and rank individuals in other groups (the $k$-partite algorithm), one where a committee is selected, and individuals are ranked recursively by larger and larger sets including the committee (the committee algorithm) and one where individuals are partitioned into two groups and rank individuals in the other group (the naive bipartite algorithm). \citet*{kahng2018ranking} provide bounds on the errors between the output of the algorithms and the ranking obtained by a deterministic social welfare function, and reports on experiments comparing the three algorithms.

Our model departs from the models of peer selection in a number of ways. First, we consider {\em ordinal} rather than cardinal information as inputs to the mechanism. In our model, individuals do not assign grades to other individuals, but can only make bilateral comparisons. Second, as in  \citet*{kahng2018ranking} but in contrast to the rest of the peer selection literature, we consider as output a {\em complete ranking} of individuals rather than a coarse ranking into two sets of acceptable and non acceptable candidates. (However, in Section \ref{section:05}, we also consider coarser rankings as a possible extension of our model.) Third, because dominant-strategy mechanisms do not exist, we weaken the incentive requirement to ex post implementation, thereby obtaining positive results which differ from the results obtained in the peer-selection literature. Fourth, and most importantly, we do not assume a specific assignment of proposals to reviewers, but consider an arbitrary network of observations captured by a social network. Our main objective is then to characterize those social networks (or structures of observability) for which mechanisms satisfying desirable properties can be constructed.

The mechanism we consider is also more distantly connected to the literature on peer prediction mechanisms initiated by \citet*{miller2005eliciting}. In a peer-prediction mechanism, there exists an unknown state of the world and individuals receive independent signals over the true state. Each individual is asked to report on his type, but receives a payment which depends on the report of another randomly chosen individual. \citet*{miller2005eliciting} show that the peer-prediction method is Bayesian incentive compatible and individually rational when individuals share a common prior. \citet*{prelec2004bayesian} extends the analysis when the planner does not know the common prior, and defines the ``Bayesian truth serum" as a mechanism where individuals receive a payment which depends on their report and on their prediction of the report of all other individuals. (The Bayesian truth serum mechanism is used by \citet*{hussam2017targeting} to elicit information about entrepreneurial abilities.) \citet*{prelec2004bayesian} shows that the Bayesian truth serum mechanism is Bayesian incentive compatible when the number of individuals becomes large. \citet*{witkowski2012robust} propose a robust peer-prediction mechanism, which does not rely on individuals sharing a common prior and, as opposed to the Bayesian truth serum,  is Bayesian incentive-compatible for small populations.

The setting we consider differs from the setting of peer prediction mechanisms in several ways. First, we assume that information is local rather than global so that individuals do not receive a signal on a common state. Second, we assume that individuals are only able to perform bilateral comparisons and cannot explicitly report cardinal types. Finally, our analysis emphasizes the importance of the network of observability rather than  scoring rules using the report of other individuals to elicit information in the community.

We also note that there exists an emerging literature in computer science on peer-grading systems.\footnote{See the website of a graduate seminar on peer-grading systems given by Jason Hartline at Northwestern University in 2017, https://sites.northwestern.edu/hartline/eecs-497-peer-grading/  for an introduction to the literature.} \citet*{de2014crowdgrader} and \citet*{wright2015mechanical} describe softwares used to implement peer-grading. \citet*{shah2013case} and \citet*{raman2014methods} argue in favor of ordinal peer-grading systems and discuss methods of converting ordinal comparisons into cardinal scales. Finally, closer to our paper, \citet*{karger2014budget} analyzes the assignment of tasks to individuals (or reviews to reviewers) under a fixed allocation of resources. They consider the problem of a planner who wants to minimize the number of reviews to reach a given reliability target, provide an adaptive algorithm and study its performance. Their analysis however does not take into account the individuals' incentives to lie or manipulate their reports.

The paper which is probably the most closely connected to ours is a recent paper by \citet*{baumann2017identifying} which analyzes network structures for which it is possible to identify the individual with the highest characteristic. \citet*{baumann2017identifying} constructs a specific multitier mechanism identifying the top individual from the reports of his friends. The mechanism admits multiple equilibria, but there are some social network architectures (e.g., the star) for which all equilibria result in the identification of the top individual. Our paper differs from Baumann's, however, in many dimensions. First, we consider an ordinal rather than a cardinal setting, giving rise to the possibility of incompleteness of the social ranking. Second, we assume that the objective of the planner is to rank all individuals rather than identify the top individual. Third, we do not assume an exogenous bound on the way in which individuals can misreport, in contrast to \citet*{baumann2017identifying}, in which this exogenous bound plays a crucial role in the construction of equilibria.

%=============================================================

\section{Model}\label{section:01}

%-------------------------------------------------------------

\subsection{Individuals and communities}

We consider a community  $N$ of $n$ individuals indexed by $i=1,2,...,n$. Each individual $i$ has a unidimensional characteristic $\theta_i$. Examples of $\theta_i$ include need, aptitude for a job, or quality of a project. We suppose that individuals cannot provide a cardinal value for the characteristic $\theta_i$. Either the characteristic cannot be measured precisely, or individuals do not have the ability or the language to quantify $\theta_i$.  Instead, we assume that individuals  possess {\em ordinal} information and are able to compare the characteristics of two individuals.

Members of the community are linked by a connected, undirected graph $g$. The social network $g$ is common knowledge among the individuals and the planner. The characteristic of individual $i$, $\theta_i$, can be observed by individual $i$ and by all his direct friends in the social network $g$.  For any individual $i$ and any pair of individuals $(j,k)$ that individual $i$ can observe, we  let $t_{jk}^i=1$ if individual $i$ observes that $\theta_j> \theta_k$, and $t_{jk}^i=-1$ if individual $i$ observes that $\theta_j < \theta_k$. The ordinal comparison is assumed to be perfect: individual $i$ always perfectly observes whether individual $j$'s characteristic is higher than that of individual $k$. We ignore situations in which the two characteristics are equal.

Individual $i$'s information (and type) can thus be summarized by a matrix $T^i = [t^i_{jk}]$, where $t^i_{jk} \in \{-1,0,1\}$ and $t^i_{jk} \neq 0$ if and only if $i$ observes the comparison between $j$ and $k$. If $i$ is one of the two individuals involved in the comparison (either $i=j$ or $i=k$),  we call the comparison $t^i_{jk}$ a {\em self-comparison}. Otherwise, if $i$ observes two different friends $j$ and $k$, $i \neq j, i \neq k$ and  $g_{ij}g_{ik} = 1$,  we call the comparison $t^i_{jk}$ a {\em friend-based comparison}.

The vector ${\bf T} = (T^1,..,T^n)$ describes the information possessed by the community on the ranking of the characteristics of all the individuals. Obviously, because individual observations are perfectly correlated, individual types $T^i$ and $T^j$ will be correlated if there exists a pair of individuals $(k,l)$ such that $t^i_{kl} \neq 0$ and $t^j_{kl} \neq 0$. Hence, if the planner could construct a punishment for contradictory reports, as in \citet*{cremer1985optimal}, she would be able to induce the individuals to report their true type. However, we rule out arbitrary punishments.

The information contained in the vector ${\bf T } = (T^1,..,T^n)$ results in a partial ranking of the characteristics of the individuals, which we denote by $\succ$. We let $i \succ_{\bf T} j$ if the information contained in ${\bf T}$ allows us to conclude that $\theta_i > \theta_j$. Formally, $i \succ_{\bf T} j$ if and only if either there exists $k$ such that $t^k_{ij} =1$ or there exists a sequence of individuals, $i^0=i,...,i^M =j$, $k^0, ...,k^{M-1}$ such that $t_{i^{m} i^{m+1}}^{k^m} =1$ for all $m=0,...,M-1$.

For a fixed social network $g$, the information contained in the vector ${\bf T} = (T^1,..,T^n)$ may not be the same for different realizations of $(\theta^1,..,\theta^n)$. This is due to the fact that (i) new comparisons can be obtained by transitivity but (ii) the transitive closure of an order relation depends on the initial order relation. To illustrate this point, consider four individuals $i=1,2,3,4$ organized in a line as in Figure \ref{figline}

%-------------------------------------------------------------
% Figure line social network on 4 nodes
% Label figline
\begin{figure}[ht]
\centering
\begin{tikzpicture}
 \SetVertexNormal[Shape = circle]
 \SetUpEdge[color = black]
    \Vertex[x=-2, y=0]{$1$}
    \Vertex[x=0, y=0]{$2$}
    \Vertex[x=2, y=0]{$3$}
    \Vertex[x=4, y=0]{$4$}
 \Edges($1$,$2$,$3$,$4$)
\end{tikzpicture}
\caption{A line of four individuals}
\label{figline}
\end{figure}
%-------------------------------------------------------------

If $\theta_1 < \theta_2 < \theta_3 < \theta_4$, then given that  $t^1_{12}=t^2_{12}=-1, t^2_{23} = t^3_{23}=-1, t^3_{34}=t^4_{34}=-1, t^2_{13}=-1$,  and $t^3_{24}=-1$, the comparisons result in a complete ranking $1 \prec  2 \prec 3 \prec 4$. However, for other possible realizations of $(\theta_1, \theta_2, \theta_3, \theta_4)$, the ranking generated by the types ${\bf T}$ may be incomplete. For example,  if $\theta_1 < \theta_4 < \theta_2 < \theta_3$, we obtain $1 \prec 2 \prec 3$ and $4 \prec 2 \prec 3$, but $1$ and $4$ cannot be compared.

A social network $g$ is called {\em completely informative} if, for any realization of the characteristics $(\theta_1, .., \theta_n)$, the information contained in ${\bf T}$ results in a complete ranking of the members of the community. The following lemma characterizes completely informative social networks.

\begin{lemma} \label{lemmacomp} A social network $g$ is completely informative if and only if, for any pair of individuals $(i,j)$ either $g_{ij}=1$ or there exists an individual $k$ such that $g_{ik} g_{jk}=1$.
\end{lemma}

A social network is completely informative if and only if {\em every pair of individuals} can be compared either by self-comparisons or by friend-based comparisons.

%-----------------------------------------------------------------

\subsection{Planner and mechanism design}

The objective of the planner is to construct a ranking of individuals according to the characteristic $\theta_i$. For example, a charity wishes to rank potential beneficiaries by need, an employer wants to rank workers according to their ability, a bank wants to rank projects according to their profitability. We let $\rho$ denote the complete order chosen by the planner. The set of all complete orders is denoted by ${\cal P}$. The rank of individual $i$ is denoted by $\rho_i$.

The planner wishes to construct a ranking as close as possible to the true ranking of the characteristic $\theta_i$. We do not specify the preferences of the planner. In the ordinal setting that we consider, different measures of distances between rankings can be constructed. Instead of describing explicitly the loss function associated with differences in rankings, we focus attention on efficient mechanisms. Efficiency requires that the ranking $\rho$ coincides with the ranking generated by ${\bf T}$ for any pair of individuals $(i,j)$ who can be compared under ${\bf T}$.

Individuals care only about their rank $\rho_i$ and have strict preferences over  $\rho_i$. By convention, individuals prefer higher ranks. Hence, $\rho_i$ is preferred to $\rho'_i$ if and only if $\rho_i > \rho'_i$. The worst rank, from the perspective of the individuals, is $\rho_i=1$. We use this convention as the worst rank will be important for incentive compatibility. We assume that there are no externalities in the community, and thus, individuals do not derive any reward from high rankings of friends or low rankings of foes.

A {\em direct mechanism} associates to any vector of reported matrices ${\bf T} \in {\cal T}^n$ a complete ranking $\rho \in {\cal P}$. We impose the following two conditions on the mechanism:

\noindent{\em Ex post incentive compatibility. } For any individual $i$, for any vector of types ${\bf T} = (T^i, T^{-i})$, any type $T^{'i} $, the following holds
\[ \rho_i ({\bf T}) \geq  \rho_i(T^{'i}, T^{-i}). \]

\noindent {\em Ex post efficiency.} For any vector of types ${\bf T}$, and for any pair of individuals $i$ and $j$, the following holds
\[ \mbox{if  }  i \succ_{\bf T} j, \mbox{ then } \rho_i({\bf T}) > \rho_j({\bf T}).\]

We focus on ex post implementation for two reasons. First, because we consider an ordinal setting, we select a robust implementation concept which does not depend on the distribution of types. Second, as we show in section \ref{section:05}, the alternative robust implementation concept---dominant-strategy implementation---is too strong for our setting.

Ex post efficiency requires that the planner's ranking coincide with the true ranking of characteristics in a weak sense. Whenever two individuals $i$ and $j$ can be ranked using the information contained in ${\bf T}$, the ranking $\rho_i$ must be consistent with the ranking between $i$ and $j$. As the order relation induced by ${\bf T}$, $\succ_{\bf T}$, may be incomplete, the requirement may be very weak. The ranking $\rho$ must be a completion of the ranking $\succ_{\bf T}$. If $\succ_{\bf T}$ is a very small subset of $N^2$, the ranking $\rho$ may end up being very different from the true ranking of the characteristic $\theta_i$. However as the true ranking of characteristics cannot be constructed using the local information from the social network, the difference between $\rho$ and the true ranking should not be a matter of concern, since the planner chooses an efficient mechanism given the information available to the community.

%=============================================================

\section{Completely informative rankings} \label{section:02}

%-------------------------------------------------------------

We first analyze conditions under which an ex post incentive-compatible and efficient mechanism can be constructed when the information available in the community always results in a complete ranking. By Lemma \ref{lemmacomp}, all pairs of individuals must either be directly connected, or observed by a third individual. The next theorem shows that for an ex post incentive-compatible and efficient mechanism to exist, all pairs of individuals must be observed by a third individual.

\begin{theorem} \label{theocomp} Suppose that the social network $g$ is completely informative. An ex post incentive-compatible and efficient mechanism exists  if and only if, for all  pair of individuals $(i,j)$,  there exists a third individual $k$ who observes both $i$ and $j$, i.e., $g_{ik}g_{jk}=1$.
\end{theorem}

Theorem \ref{theocomp} shows that an ex post incentive-compatible and efficient mechanism exists in completely informative communities  if and only if every pair of individuals $(i,j)$ has a common friend $k$. Self-comparisons cannot be used if they are not supported by the report of a third party.

The intuition underlying Theorem \ref{theocomp} is easy to grasp. If the comparison between $\theta_i$ and $\theta_j$ can be reported only by $i$ and $j$, in an ex post efficient mechanism, one of them has an incentive to lie. Consider a ranking which places $i$ and $j$ as the two individuals with the lowest characteristics in the community. If both announce that $\theta_i$ is smaller than $\theta_j$, then $\rho_i=1, \rho_j=2$. Similarly, if both announce that $\theta_j$ is smaller than  $\theta_i$, then $\rho_j=1, \rho_i=2$. But by incentive compatibility, neither of the individuals can improve his rank by changing his report on $t_{ij}$. Hence $i$ must still be ranked at position 1 when he announces $\theta_i > \theta_j$ and $j$ announces $\theta_i < \theta_j$, and similarly individual $j$ must still be ranked at position 1 when he announces $\theta_j > \theta_i$ and individual $i$ announces $\theta_i > \theta_j$. As two individuals cannot occupy the same position in the ranking, this contradiction shows that there is no ex post incentive-compatible and efficient mechanism relying on self-comparisons. Notice that this impossibility result stems from the fact that the planner has a very small number of outcomes at her disposal.  If she could  impose any arbitrary punishment (for example by excluding all individuals  who provide inconsistent reports), she could implement an ex post efficient mechanism in dominant strategies, as in  \citet*{cremer1985optimal}, for any network architecture.

The construction of an ex post incentive-compatible and efficient mechanism when all links are supported requires more work. Consider first the simple case where a comparison between $i$ and $j$  is observed by at least three individuals. The mechanism disregards the report of any individual who deviates from the reports of all other individuals. Hence no individual can unilaterally change the outcome of the mechanism when all other individuals report the truth. Next consider the more complex situation where $i$ and $j$ are not connected and  the comparison between $i$ and $j$ is dictated by a third party, a common friend $k$. We show that a  change in his report cannot improve the rank of $k$ given that all other individuals tell the truth and that the social network is completely informative.  If the change in report creates an inconsistency in the ranking, the mechanism guarantees that  the planner can detect if a single individual has cheated. If this is the case, the mechanism punishes the defector by ranking him at the worst position in the ranking. If the change in report does not create a violation in transitivity, because the social network is completely informative, the rank of individual $k$ is fully determined by the reports of other individuals in the community. The rank of individual $k$ is fixed and no change in report can improve the position of individual $k$ in the ranking. This friend-based ranking mechanism is ex post incentive-compatible and efficient.

Theorem \ref{theocomp} characterizes communities for which friend-based ranking mechanisms can be constructed. Clearly the complete network satisfies the conditions. However, the condition is also satisfied by many other social networks, which are less dense than the complete network. Our next result characterizes the {\em sparsest} networks for which the condition of Theorem \ref{theocomp} holds. This characterization is based on the ``friendship theorem" of  \citet*{erdos1966problem}.

\begin{theorem} (The friendship theorem) If $g$ is a graph of $n$ nodes in which any two nodes $i$ and $j$ have exactly one friend in common, then there exists $m$ such that $n=2m+1$  and $g$ contains $m$ triangles which are connected at a common node.
\end{theorem}

The friendship theorem, initially stated and proved in \citet*{erdos1966problem}, asserts that in any community where every pair of individuals has exactly one friend in common, the number of individuals (the order of the graph) needs to be odd and one individual is friends with everyone and is the common friend of all other individuals.\footnote{Different proofs of the friendship theorem have been proposed, often using complex combinatorial arguments \citep{wilf1971friendship,longyear1972friendship,huneke2002friendship}.} The ``friendship graph" is illustrated in Figure \ref{fig:windmill} for $n=7$. For obvious reasons, it is also called the windmill graph. The friendship graph has exactly $3m$ edges. Our next theorem shows that this is actually the smallest number of edges for which a completely informative mechanism can be constructed when $n$ is odd. When $n$ is even, the graph which minimizes the number of edges is a variation of the friendship graph, where one of the sails of the windmill contains three vertices, as illustrated in Figure \ref{fig:windmill} for $n=8$.

%-------------------------------------------------------------
% Friendship graph on 7 nodes
% Figurewindmill on 8 nodes
\begin{figure}[ht!]
\centering
\begin{subfigure}[b]{0.4\textwidth}
\centering
\begin{tikzpicture}
   \GraphInit[vstyle=Hasse]
   \SetUpEdge[color = black]
      \Vertex[x=2, y=2]{1}
      \Vertex[x=0, y=2]{2}
      \Vertex[x=1, y=4]{3}
      \Vertex[x=3, y=4]{4}
      \Vertex[x=4, y=2]{5}
      \Vertex[x=3, y=0]{6}
      \Vertex[x=1, y=0]{7}
   \Edges(1,2)
   \Edges(1,3)
   \Edges(1,4)
   \Edges(1,5)
   \Edges(1,6)
   \Edges(1,7)
   \Edges(2,3)
   \Edges(4,5)
   \Edges(6,7)
  \end{tikzpicture}
\caption{$n=7$. For odd number of nodes, the windmill is also called a friendship graph.}
\end{subfigure}
\quad \quad
\begin{subfigure}[b]{0.4\textwidth}
\centering
\begin{tikzpicture}
   %\GraphInit[vstyle=Classic]
   %\SetVertexNoLabel
   \GraphInit[vstyle=Hasse]
   \SetUpEdge[color = black]
      \Vertex[x=2, y=2]{1}
      \Vertex[x=0, y=2]{2}
      \Vertex[x=1, y=4]{3}
      \Vertex[x=3, y=4]{4}
      \Vertex[x=4, y=2]{5}
      \Vertex[x=3.5, y=0]{6}
      \Vertex[x=2, y=0]{7}
      \Vertex[x=0.5, y=0]{8}
   \Edges(1,2)
   \Edges(1,3)
   \Edges(1,4)
   \Edges(1,5)
   \Edges(1,6)
   \Edges(1,7)
   \Edges(1,8)
   \Edges(2,3)
   \Edges(4,5)
   \Edges(6,7,8)
  \end{tikzpicture}
\caption{$n=8$. For even number of nodes the windmill is modified and one sail has three nodes.}
\end{subfigure}
\caption{Windmill graphs}
\label{fig:windmill}
\end{figure}

%-------------------------------------------------------------

\begin{theorem} \label{theowindmill} Suppose that $n \geq 3$. Let $g$ be a social network for which friend-based ranking generates a complete ranking. Then $g$ must contain at least $\frac{3n}{2}-1$ links if $n$ is even and $\frac{3(n-1)}{2}$ links is $n$ is odd. If $n$ is odd, the unique sparsest network architecture is the friendship network. If $n$ is even, the unique sparsest network architecture is a modified windmill network where one of the sails contains three nodes $i,j,k$ such that $i,j$ and $k$ are connected to the hub, $i$ is connected to $j$ and $j$ is connected to $k$.
\end{theorem}

Theorem \ref{theowindmill} establishes a lower bound on the number of edges needed to obtain a complete ranking of the community. It also identifies the unique network architecture which reaches this lower bound: a windmill network where one of the nodes, the hub, connects all other nodes which form pairs.\footnote{The proof of the theorem is very different from known proofs of the Friendship Theorem, mostly because we focus attention on the minimization of the number of edges rather than on the construction of a graph where any intersection of neighborhoods is a singleton.} This network architecture implies a very unequal distribution of degrees. The hub is connected to all nodes, whereas the remaining nodes have degree two or three. If individuals have a limited capacity to compare other individuals, the windmill network cannot be used, and one needs to resort to other more symmetric network architectures involving a larger number of links. An exact characterization of the minimal degree of a regular network for which all links can be supported remains an open question in graph theory.\footnote{A family of regular graphs, called the rook graphs, satisfy the property. For an integer $m \geq 2$, rook graphs are regular graphs of degree $2(m-1)$ among $m^2$ nodes, and have the property that any two connected nodes have $m-2$ nodes in common and every pair of unconnected nodes has two common friends. See \citet*{brouwer2011spectra} for more details on rook graphs.}

%-------------------------------------------------------------

\section{Incomplete rankings} \label{section:02b}

We now consider communities which are not completely informative. There are realizations of the type ${\bf T}$ for which individuals collectively cannot construct a complete ranking. Some individuals cannot be ranked, and we denote by $i \bowtie_{\bf T} j$  the fact that $i$ and $j$ cannot be compared using the information contained in ${\bf T}$. Note however that  the mechanism $\rho$ defines a complete ranking of all individuals in the community. It must thus choose an arbitrary ranking between incomparable individuals at ${\bf T}$. We first show that Theorem \ref{theocomp} can be generalized to communities which are not completely informative.

\begin{theorem} \label{theocomp2} An ex post incentive-compatible and efficient mechanism exists  if and only if, for every pair of linked individuals $(i,j)$ (i.e. $g_{ij}=1$),  there exists a third individual $k$ who observes both $i$ and $j$, i.e., $g_{ik}g_{jk}=1$.
\end{theorem}

%-------------------------------------------------------------
% Figure g supported
% Label: figsupport
\begin{figure}[ht]
\centering
\begin{tikzpicture}
 \GraphInit[vstyle=Hasse]
 \SetUpEdge[color = black]
    \Vertex[x=0, y=0]{1}
    \Vertex[x=1, y=1]{2}
    \Vertex[x=1, y=-1]{3}
    \Vertex[x=2, y=0]{4}
    \Vertex[x=3, y=1]{5}
    \Vertex[x=4, y=0]{6}
    \Vertex[x=5, y=1]{7}
    \Vertex[x=5, y=-1]{8}
    \Vertex[x=6, y=0]{9}
 \Edges(1,2,4,5,6,7,9)
 \Edges(1,3,4,6,8,9)
 \Edges(1,4)
 \Edges(6,9)
\end{tikzpicture}
\caption{A supported social network $g$}
\label{figsupport}
\end{figure}
%-------------------------------------------------------------

Theorem \ref{theocomp2} identifies network structures under which the planner can extract all local information from the community, for any realization of characteristics. All friends must have a friend in common and the social network  is thus formed of a collection of triangles. This is similar to the structure of communities identified in \citet*{jackson2012social} where any social link is ``supported" by links to a common friend.  Figure \ref{figsupport} illustrates one of these networks. Notice that some comparisons are supported as links within the triangles whereas other comparisons are supported as links across triangles. Links across triangles do not play a role in the \citet*{jackson2012social} favor exchange context.

The proof of Theorem extends the arguments of Theorem \ref{theocomp}. For the necessity part, we show that the planner cannot construct an ex post incentive-compatible and efficient mechanism when two individuals provide self-comparisons. The argument is more complex than in completely informative societies because we need to guarantee that whenever two players are ranked last when they provide consistent assessments, they remain ranked last when they provide inconsistent reports. For the sufficiency part, we construct the mechanism as in Theorem \ref{theocomp}, but the proof that a third individual $k$ cannot improve his ranking by changing his report on $i$ and $j$ is much more involved. We first  need to extend the argument that the planner can detect when a single individual has created an intransitivity in the ranking. But, more importantly, we check that, when the reports result in a transitive ranking, individual $k$ cannot improve his ranking by changing his report, because any pair of incomparable individuals must either be both ranked above or both ranked below $k$.
%=============================================================

\section{Rankings using only friend-based comparisons} \label{section:03}

%-------------------------------------------------------------

\subsection{Comparison networks}

As the necessity arguments in the proofs of Theorem \ref{theocomp} and \ref{theocomp2} show, self-comparisons cannot be used to extract information.  In this section, we consider the following problem. Suppose that the planner discards self-comparisons. When is it possible to use all remaining local comparisons in the network?

To answer this question, we first modify the type of individual $i$, $T^i$ by removing any comparison $t_{ij}^i$ which is not supported by a third individual, i.e., we let $t_{ij}^i = 0$ if there exists no $k \neq i,j$ such that $t^k_{ij} \neq 0$. Second, we construct a {\em comparison network $h$} which captures all comparisons that can be obtained using friend-based comparisons. Formally, we let $h_{ij}=1$ if and only if there exists $k$ such that $g_{ik}g_{jk} = 1$. The network $h$ collects all pairs of individuals which can be compared by a third individual. It differs from the social network $g$ in two ways: (i) pairs of individuals which are linked in $g$ but do not have a common friend appear in $g$ but not in $h$, (ii) pairs of individuals which are not directly linked in $g$ but have a common friend appear in $h$ but not in $g$.  Figure \ref{figcompnet} illustrates a social network $g$ and the corresponding comparison network $h$.

%-------------------------------------------------------------
% Figure example g with corresponding h
% Label: figcompnet
\begin{figure}[ht]
\centering
\begin{subfigure}[b]{\textwidth}
\centering
\begin{tikzpicture}
 \GraphInit[vstyle=Hasse]
 \SetUpEdge[color = black]
    \Vertex[x=0, y=0]{1}
    \Vertex[x=6, y=0]{2}
    \Vertex[x=2, y=1]{3}
    \Vertex[x=4, y=1]{4}
    \Vertex[x=6, y=2]{5}
    \Vertex[x=0, y=2]{6}
    \Vertex[x=3, y=2]{7}
 \Edges(5,2,4,3,6,7)
 \Edges(6,1,3)
 \Edges(5,4)
\end{tikzpicture}
\caption{Social network $g$}
\end{subfigure}
\newline
\begin{subfigure}[b]{\textwidth}
\centering
\begin{tikzpicture}
 \GraphInit[vstyle=Hasse]
 \SetUpEdge[color = black]
    \Vertex[x=0, y=0]{1}
    \Vertex[x=6, y=0]{2}
    \Vertex[x=2, y=1]{3}
    \Vertex[x=4, y=1]{4}
    \Vertex[x=6, y=2]{5}
    \Vertex[x=0, y=2]{6}
    \Vertex[x=3, y=2]{7}
 \Edges(5,2,4,6,3,1)
 \Edges(6,1,4,5,4)
 \Edges(3,7)
 \Edges (3,5)
 \Edges (3,2)
 \tikzset{EdgeStyle/.style={bend left}}
 \Edges(1,7)
\end{tikzpicture}
\caption{Comparison network $h$}
\end{subfigure}
\caption{Social and comparison networks}
\label{figcompnet}
\end{figure}
%-------------------------------------------------------------

The comparison network $h$ is the  set of bilateral comparisons that the planner can guarantee by excluding self-comparisons. The planner complements the comparisons contained in $h$ by taking their transitive closure.

We say that a mechanism \textit{supports} the comparison network $h$ if and only if the social planner's ranking coincides with the comparisons in $h$.  Formally, let $T^i_h$ be the restriction of $T^i$ to pairs $(j,k)$ which are connected in $h$. A direct mechanism associating to every ${\bf T}_h$ a complete ranking $\rho$ supports $h$ if and only if it satisfies ex-post incentive compatibility and ex-post efficiency for any pair $(i,j)$ in $h$. Notice that in particular that the mechanism discards all information about pairs which do not belong to $h$ and faces no constraint in the ranking of these pairs.

Clearly, the empty network is always supported by any mechanism.  If the conditions of Theorem \ref{theocomp} hold, there exists a mechanism supporting the complete network. When the conditions fail, our objective is to construct the maximal number of bilateral comparisons, namely to identify maximal comparison networks which can be supported by some mechanism.

Consider as an example of a social network $g$ the four-individual line of Figure \ref{figline}. The comparison network $h$ associated to $g$ consists of the two linked pairs $(1,3)$ and $(2,4)$. A mechanism supporting $h$ only uses $2$'s ranking of $1$ and $3$ and $3$'s ranking of $1$ and $4$. It must satisfy ex-post incentive compatibility and ex-post efficiency over the pairs $(1,3)$ and $(2,4)$. All other bilateral rankings are unconstrained and can be chosen arbitrarily. Consider the mechanism which follows $2$'s ranking between $1$ and $3$, $3$'s ranking between $1$ and $2$ and always ranks $1$ and $3$ above $2$ and $4$. This mechanism supports $h$ as individuals $2$ and $3$ cannot change their ranks by their announcement, and the mechanism follows the ranking prescribed by $2$ and $3$ over the pairs $(1,3)$ and $(2,4)$.

%-----------------------------------------------------------------

\subsection{Bipartite social networks}

We first provide a characterization of these social networks $g$ which generate connected comparison networks $h$. Our characterization is in terms of bipartite social networks. A social network is \textit{bipartite} if the individuals can be divided into two disjoint sets $A$ and $B$ such that individuals in $A$ only link to individuals in $B$ and individuals in $B$ only link to individuals in $A$.

\begin{proposition} Suppose that $n \geq 3$.  The comparison network $h$ is connected if and only if  $g$ is not bipartite.
\label{proconn}
\end{proposition}

Proposition \ref{proconn} establishes that the network $h$ is connected if and only if the social network $g$ is not bipartite. If the network $g$ is not bipartite, we construct a path connecting any pair of individuals $i$ and $j$ in the comparison network. If the network $g$ is bipartite, and the nodes partitioned into the two sets $A$ and $B$, the comparison network $h$ is disconnected into two components: individuals in $A$ rank individuals in $B$ and individuals in $B$ rank individuals in $A$.  Individuals can be ranked inside the two sets $A$ and $B$ but rankings of individuals across the two sets must be arbitrary. Notice however that an individual in $A$ cannot improve his ranking by lying about the ranking of individuals in $B$. Hence, when the social network $g$ is bipartite (and the comparison network $h$ disconnected), it is easy to construct a mechanism satisfying ex post incentive compatibility and efficiency.

\begin{proposition} Suppose that the social network $g$ is bipartite with two sets of nodes $A$ and $B$. Then there exists a mechanism supporting the comparison network $h$ on its two components $A$ and $B$.
\label{probipartite}
\end{proposition}

Proposition \ref{probipartite} characterizes one situation where the planner can elicit information about  comparisons: when the set of individuals in the community can be partitioned into two subsets where members of one subset observe members of the other subset. For example, one could survey separately men and women and ask men about the characteristics of women and women about the characteristics of men. However, this design would not allow the planner to obtain information about the ranking of individuals across the two sets. The mechanism completes the partial ranking by an arbitrary ranking across individuals in the two sets, possibly resulting in a final ranking which is very different from the true ranking.

We observe that the partition of the set of nodes into groups which rank each other is the basis of most algorithms proposed in the computer science literature on peer selection. The partition mechanism allows for strategy-proof implementation, a stronger incentive compatibility requirement than ex post incentive compatibility.

%-----------------------------------------------------------------

\subsection{Connected comparison networks}

We now consider communities for which the comparison network $h$ is connected but not complete. From Theorem \ref{theocomp2}, we know that if the graph $g$ only contains triangles, then there exists an ex-post efficient and incentive compatible mechanism supporting the  comparison network $h$. What if the network $g$ in neither bipartite nor only formed of triangles? The following simple example shows that it may be impossible to find an ex-post efficient and incentive compatible mechanism supporting the comparison network $h$.

\begin{example} Let $n=4$. Individuals $i,j,k$ are connected in a triangle and individual $l$ is connected to $i$. \label{exinexist} \end{example}

%-------------------------------------------------------------
% Figure kite counter example
% Label: figinexist
\begin{figure}[ht]
\centering
\begin{tikzpicture}
 \SetVertexNormal[Shape = circle]
 \SetUpEdge[color = black]
    \Vertex[x=1, y=2]{$i$}
    \Vertex[x=0, y=0]{$j$}
    \Vertex[x=2, y=0]{$k$}
    \Vertex[x=4, y=0]{$l$}
 \Edges($i$,$j$,$k$,$i$)
 \Edges($i$,$l$)
\end{tikzpicture}
\caption{A social network $g$ where a mechanism does not exist}
\label{figinexist}
\end{figure}
%-------------------------------------------------------------

In this example, the links $(i,j), (i,k),(j,k)$ are supported, but the link $(i,l)$ is not supported. Consider a realization of the characteristics such that $\theta_ l > \theta_j > \theta_i > \theta_k$. If individual $i$ announces $\theta_l > \theta_ j > \theta_k$, by ex post efficiency, the planner constructs the rankings $k,i,j,l$ and the rank of individual $i$ must be equal to $2$. If on the other hand individual $i$ announces $\theta_j > \theta_k > \theta_l$, the planner constructs the ranking $l,k,i,j$ and the ranking of individual $i$ is now equal to $3$. Hence individual $i$ has an incentive to lie and announce $\theta_j > \theta_k > \theta_l$.

In Example \ref{exinexist}, the planner's ranking of $i$ depends on his announcement on the rankings $(j,l)$ and $(k,l)$. Given that $\theta_j > \theta_i > \theta_k$, and that $(i,j,k)$ form a  triangle, the planner must rank $i$ between $j$ and $k$. Hence she cannot rank all three individuals $j,k$ and $l$ on the same side of individual $i$, as in the mechanism constructed in the proof of Theorem \ref{theocomp2}. But then, the announcement of individual $i$ on $(j,k,l)$, by changing the rank of $l$ with respect to $j$ and $k$, will also affect the ranking of individual $i$. Because individual $i$ can manipulate his rank by his announcements on the unsupported links $(j,k)$ and $(j,l)$, there is no mechanism satisfying ex post incentive compatibility and efficiency in this community.

%=============================================================

\section{Real-life social networks}\label{section:04}

We next illustrate our theoretical findings with data from real-life social networks collected by \citet{banerjee2013diffusion} in India and \citet{alatas2016network} in Indonesia. Our theoretical analysis highlights the importance of two network measures. In order to construct a complete ranking, Theorem \ref{theocomp} shows that every pair of individuals must be observed by a third individual. Hence, the {\em density of the comparison network}, which measures the fraction of pairs which are observed by a third individual, serves as an indicator of the possibility to construct a truthful complete ranking.  In order to construct an efficient and truthful ranking, Theorem \ref{theocomp2} asserts that every pair of friends must have a friend in common. The {\em support measure} computes the fraction of links which are supported by the presence of two additional links to a third individual, and serves as an indicator of the possibility to construct a truthful efficient ranking.\footnote{Our support measure is identical to the self-support measure defined in \citet*{jackson2012social}.}   Of course, because the condition of Theorem \ref{theocomp2} is weaker than the condition of Theorem \ref{theocomp}, the density of the comparison network is always lower than the support measure. We also study the construction of maximal comparison networks  using two different strategies suggested by Theorem \ref{theocomp2} and Proposition \ref{probipartite}: (i) elimination of links which are not parts of triangles and (ii) elimination of links to obtain a bipartite network.

The data collected by  \citet{banerjee2013diffusion} in India and \citet{alatas2016network} in Indonesia are very rich and  contain multiple independent networks: 75 villages from Karnataka, India, and 633 hamlets from three provinces in Indonesia.\footnote{The Indian dataset maps several networks for each village. We aggregate all the social networks into a {\em union} by constructing a link between two households if and only if they are connected in any of the social networks.}  We cannot directly compare the Indian networks with the Indonesian networks as different survey methods were employed in each setting. In India all households were surveyed while in Indonesia only 9 households were surveyed in each hamlet (8 randomly selected households and the hamlet head). While sampled households in Indonesia were also asked about links to out-of-sample households, there was a clear concentration of social links among the sampled households. This partly explains why social networks in Indonesia are found to have higher density and average clustering than in India. Since our theoretical results assume the network is connected, we focus on the giant component (the largest connected component) of each network.

 %--- Table
 \begin{table}[ht]
\centering
\small
\caption{Summary statistics of social networks}
\label{table:net_sumstats}
%\footnotesize
\begin{threeparttable}
\begin{tabulary}{\textwidth}{CCLCL}
\toprule
	&	\textbf{India}	& &	\textbf{Indonesia} & \\
\midrule
Networks        & 75 &   & 633 &\\
& & & &\\
& Mean & [Min, Max]& Mean & [Min, Max] \\
Households in giant component                   & 188.87 & [75, 341] & 23.6 & [2, 82]          \\
Density of social network           	& 0.05 & [0.02, 0.12] & 0.36 & [0.09, 1.00] \\
Average clustering                      & 0.26 & [0.16, 0.45] & 0.73  & [0.00, 1.00] \\
Density of comparison network           & 0.37 & [0.18, 0.62] & 0.70  & [0.00, 1.00] \\
Support                                 & 0.85 & [0.68, 0.95] & 0.95 & [0.00,1.00] \\
\bottomrule
\end{tabulary}
\begin{tablenotes}
\item Note: All statistics are calculated on the giant component. Data is sourced from \citet{banerjee2013diffusion} for India and \citet{alatas2016network} for Indonesia.
\end{tablenotes}
\end{threeparttable}
\end{table}

Table \ref{table:net_sumstats} provides summary statistics of the networks. We report the mean, minimum, and maximum for each measure.   In addition to the density of the comparison network and the support measure, we compute two common network statistics. The density of the social network measures the fraction of links formed with respect to the maximal possible number of links and average clustering measures the average fraction of links formed between friends of each individual.

%-------------------------------------------------------------
% Figure:
\begin{figure}[ht]
\centering
\begin{subfigure}{0.49\textwidth}
\includegraphics[width=\textwidth]{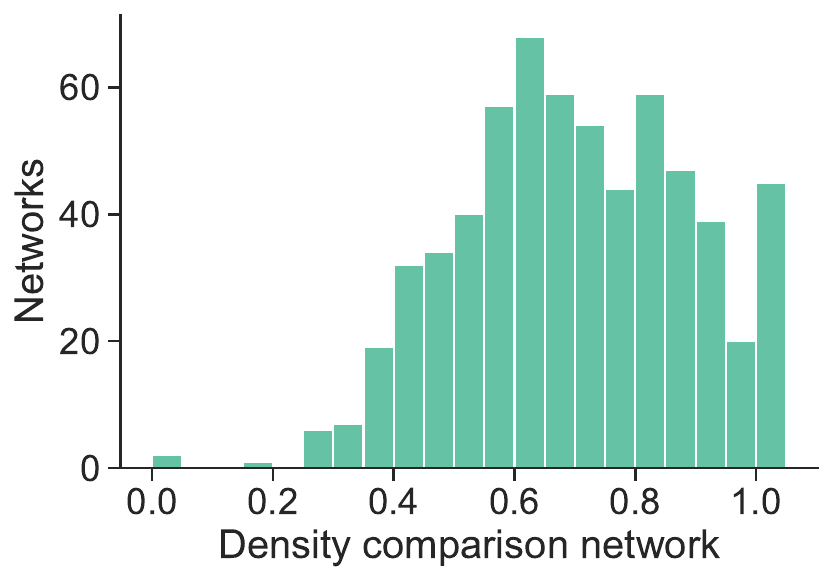}
\caption{Indonesian networks}
\end{subfigure}
\begin{subfigure}{0.49\textwidth}
\includegraphics[width=\textwidth]{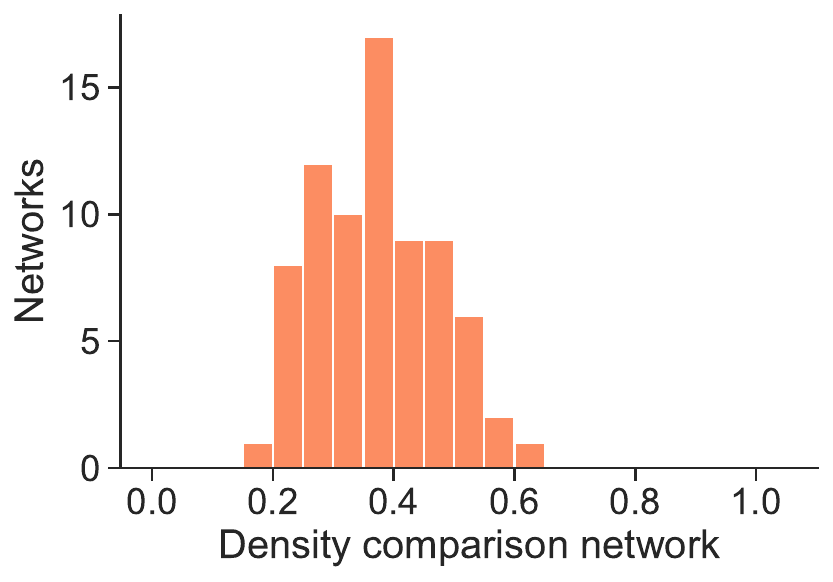}
\caption{Indian networks}
\end{subfigure}
\caption{Histograms of the density of the comparison network}
\label{fig:den_comp}
\end{figure}
%-------------------------------------------------------------

Figure \ref{fig:den_comp} provides histograms of the density of the comparison network.  For the Indonesian networks, 45 of the 633 networks have a complete comparison network. None of the larger Indian networks have a complete comparison network. As networks grow large, the number of pairs grows rapidly so that it becomes more difficult to satisfy the requirement of Theorem \ref{theocomp} and construct a complete ranking.

%-------------------------------------------------------------
% Figure:
\begin{figure}[ht]
\centering
\begin{subfigure}{0.49\textwidth}
\includegraphics[width=\textwidth]{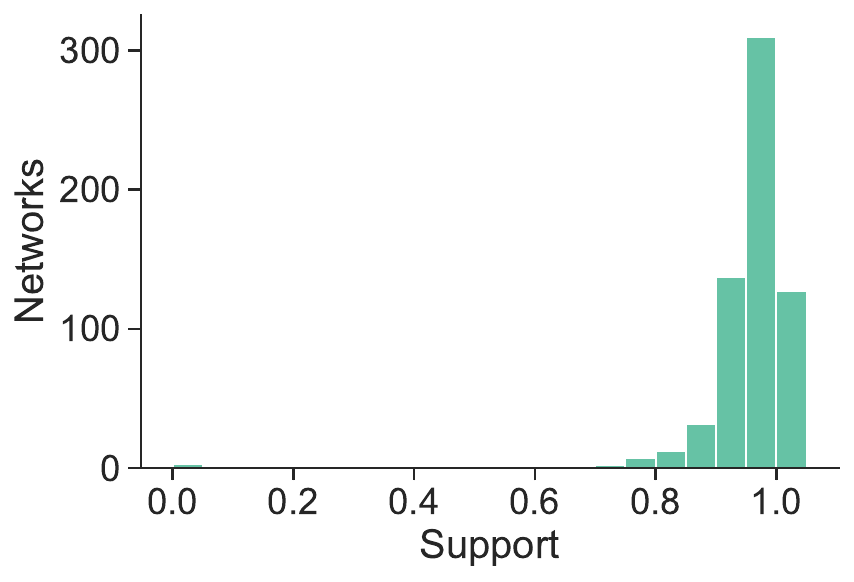}
\caption{Indonesian networks}
\end{subfigure}
\begin{subfigure}{0.49\textwidth}
\includegraphics[width=\textwidth]{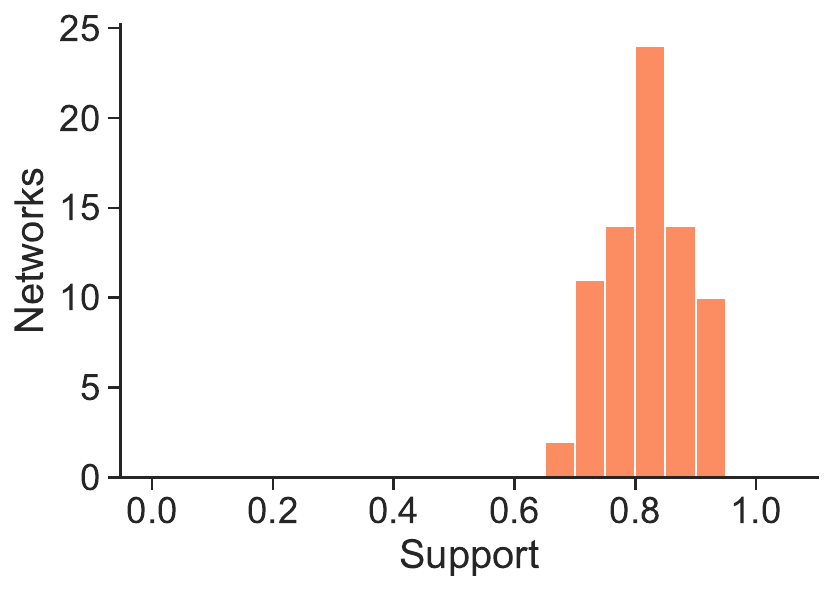}
\caption{Indian networks}
\end{subfigure}
\caption{Histograms of share of supported links}
\label{fig:support}
\end{figure}
%-------------------------------------------------------------

Figure \ref{fig:support} provides histograms of the support measure. We verify that the support measure is always higher than the density of the comparison network. For the Indonesian networks,  127 out of 633 have full support.  None of the Indian networks have full support but most have at least 80 percent of links supported.

\begin{figure}[ht]
\centering
\includegraphics[width=\textwidth]{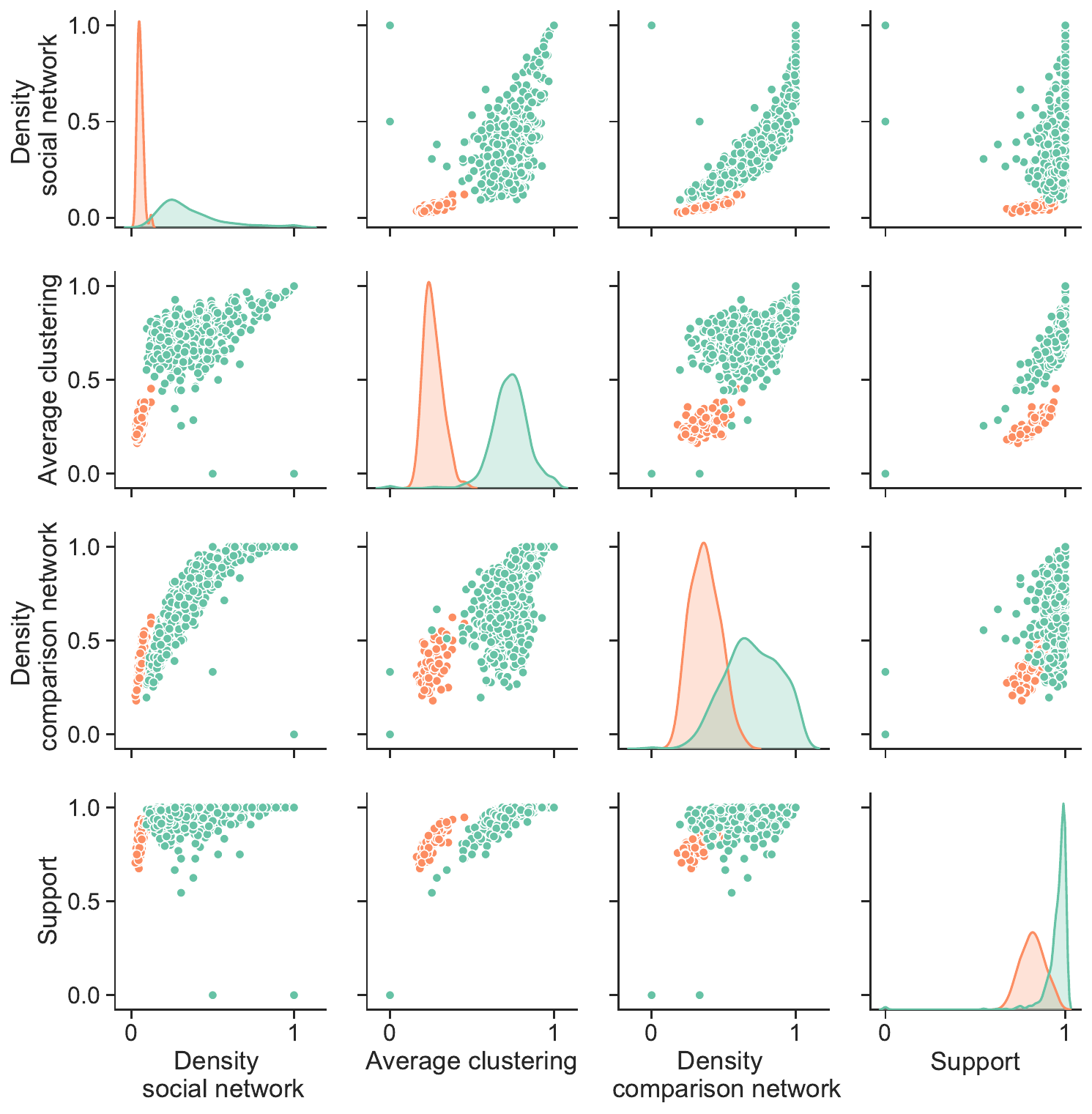}
\caption{Scatter plots of social network measures}
\fignote{Note: Social networks from India (in orange) and Indonesia (in green).}
\label{fig:pairplot}
\end{figure}

All network measures are obviously correlated. Figure \ref{fig:pairplot} plots the correlation between the four network measures we consider. The data confirms the well-known positive correlation between the density of the network and average clustering. Interestingly, density of the social network and density of the comparison network are related though a  positive but non-linear relation. While denser networks also have a denser comparison network, there exists a large fraction of social networks in Indonesia with very high density of the comparison network but variable density of the social network. The density of the comparison network is also positively correlated with average clustering, as more triangles increase the fraction of individuals who are connected through a friend. The support measure is very high in our sample. It is weakly positively correlated with average clustering---an observation that was already made by \citet*{jackson2012social} for the Indian data. The correlation between the support measure and the two other network measures (density of the social and comparison networks) appears to be weak.

%-------------------------------------------------------------
% Figure:
%-------------------------------------------------------------

%-- I slightly rewrote this paragraph to make it more readable

Most of the social networks do not meet the exact requirements of Theorem \ref{theocomp} or \ref{theocomp2}. For these networks, we  compute the largest set of comparisons that the planner can extract through an efficient and incentive compatible mechanism.  Our theoretical analysis provides the planner with two options. First, she could use a {\em support} mechanism by removing all links which are not supported  (i.e. not part of a triangle). Second, she could use a {\em bipartite} mechanism, by partitioning the set of individuals into two groups, $A$ and $B$, and have individuals in each group rank the individuals of the other group. In order to compute the optimal bipartite mechanism, we search across every possible partition to determine which partition results in the largest number of bilateral comparisons. Since this process is computationally intensive, we limit our sample to the  213 networks with 20 or less households in the giant component. All of these networks are from the Indonesian data.

\begin{figure}
    \centering
    \includegraphics[width=0.5\textwidth]{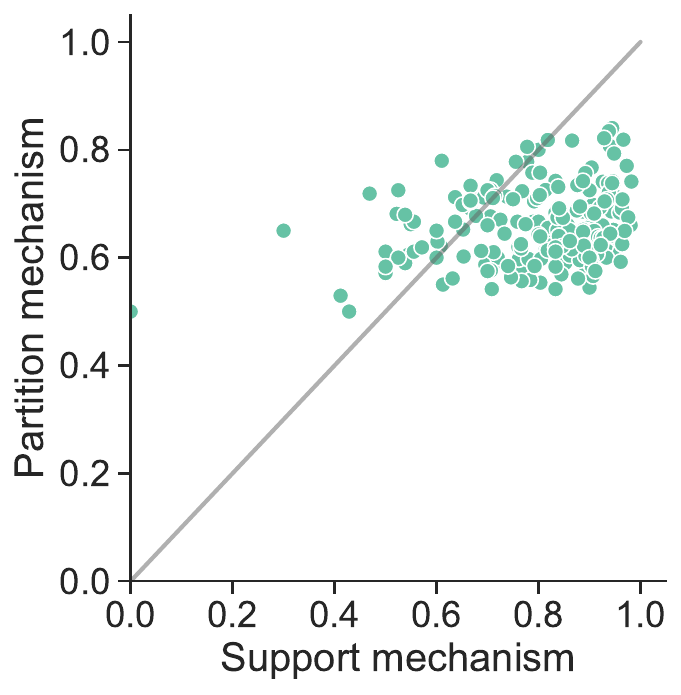}
    \caption{Share of comparisons retained by the support and partition mechanisms}
    \fignote{Note: Due to computational constraints, only networks with 20 or less households in the giant component are included in this figure. All of these small networks are from Indonesia.}
    \label{fig:mechanisms}
   \end{figure}

Figure \ref{fig:mechanisms} plots the share of comparisons retained by the optimal partition and support mechanisms for each of the 213 networks. The share is computed by dividing the number of comparisons retained through the mechanism by the total number of friend-based comparisons. The average share of comparisons retained by the support mechanism, 0.6, is larger than the average share retained by the partition mechanism,  equal to 0.5. Out of the 213 networks shown on Figure \ref{fig:mechanisms}, 162 networks are below the 45 degree line (implying that the support mechanism results in a larger share than the partition mechanism) and 51 are on or above the 45 degree line (indicating that the partition mechanism results in a larger or equal share compared to the support mechanism).  This pattern suggests that the  support mechanism provides a more effective means of extracting comparisons than the partition mechanism.

We have exploited data on real social networks in India and Indonesia as a mere illustration of  the theoretical analysis of the paper. We are aware that further full-fledged empirical tests of the model could be performed using the rich data collected in Indonesia. These data include measures of accuracy of the community ranking and individual wealth comparisons provided by 8 randomly selected households. Even though the community ranking mechanism and the mechanism studied in this paper are different, we could try and connect the accuracy of the community ranking to the architecture of the social network. In particular, we wonder whether networks with more triangles allow for better community rankings. The individual wealth comparisons could also be exploited to study the accuracy of a mechanism using (i) only self-comparisons, (ii) only friend-based comparisons or (iii) both friend-based and self-comparisons.

%=============================================================

\section{Robustness and extensions} \label{section:05}

%-------------------------------------------------------------
The analysis of friend-based ranking mechanisms relies on specific assumptions on the model. In this section, we relax some of these assumptions to test the robustness of our results.

\subsection{Dominant strategy implementation}

We first strengthen the incentive compatibility requirement to dominant-strategy implementation. The following proposition shows that dominant-strategy implementation is too strong in our setting.  The outcome set is not rich enough to permit the construction of strategy-proof mechanisms. We first recall the definition of  strategy-proofness:

\noindent{\em Strategy-proofness.} For any individual $i$, for any vector of announcements $(\hat{T}^{-i})$ and any types $T^i, T^{'i}$,
\[ \rho_i (T^i, \hat{T}^{-i}) \geq  \rho_i(T^{'i}, \hat{T}^{-i}). \]

\begin{proposition}  \label{proptriangleds} Let $g$ be a triangle. There exists no mechanism satisfying strategy-proofness and ex post efficiency.
\end{proposition}

Proposition \ref{proptriangleds} is an impossibility result, highlighting a conflict between strategy-proofness and efficiency in a very simple network architecture. As shown in the proof, the impossibility stems from the coarseness of the outcome space, which limits the power of the planner. There are only three possible outcomes corresponding to the three possible ranks. Strategy-proofness imposes a large number of constraints on the mechanism. We show, using a combinatorial argument, that if all the constraints are satisfied, two individuals must be occupying the same rank for some vectors of announcements. Hence, it is impossible to elicit truthful information in a complete network with three individuals.\footnote{The extension of this impossibility result to more than three individuals remains an open question.}

%-------------------------------------------------------------

\subsection{Coarse rankings}

We next relax the assumption that the planner chooses a complete ranking and that individuals have strict preferences over ranks. We consider a setting where the planner selects only broad indifference classes. This is the typical situation in which the planner selects a set of  recipients of the benefits of social programs, or of research funds.
If the planner only chooses broad categories, she might be able to construct ex post incentive-compatible and efficient mechanisms even if self-comparisons are not supported by a third individual. The intuition is immediate: if there exist two ``worst spots" in the ranking, the planner can punish individuals who send conflicting self-comparisons by placing  both of them on the worst spot. We formalize this intuition in the following proposition.

\begin{proposition} \label{coarse}
 There exists an ex post incentive-compatible and efficient mechanism in any completely informative community if and only if the planner can place two individuals in the worst spot, i.e.,  if and only if any individual $i$ is indifferent between $\rho(i)=1$ and $\rho(i)=2$.
\end{proposition}

Proposition \ref{coarse} thus shows that it is easier to construct ex post incentive-compatible and efficient mechanisms when the planner does not construct a complete ranking of the individuals. This observation raises new possibilities. It may be possible to construct incentive-compatible and efficient mechanisms when the planner only assigns individuals to broad categories. This also suggests that it may be possible to use local information to rank individuals into quantiles.

%-------------------------------------------------------------

\subsection{Group incentive compatibility}

We now allow for individuals to jointly deviate from truth-telling. We let individuals coordinate their reports and jointly misreport their types.

Consider a triangle with three individuals. Each individual reports on the three links. The mechanism that we constructed in Theorem \ref{theocomp} of Section \ref{section:02} assigns a ranking $\rho(i) > \rho(j)$ when at least two of the individuals report that $i$ is higher than $j$. This creates an incentive for any pair of individuals to misrepresent their types. For example, if the true ranking is $\theta_3 > \theta_2 > \theta_1$, individuals $1$ and $2$ have an incentive to misreport and announce that $2$ is higher than $1$, and $1$ is higher than $3$, so that in the end, $\rho(2)=3 > 2$ and $\rho(1) = 2> 1$.

This intuition can be exploited to show that there does not exist any mechanism satisfying ex post group-incentive compatibility and efficiency when $n=3$. We first provide a formal definition of ex post group-incentive compatibility:

\noindent{\em Ex post group incentive compatibility. } For any vector of types ${\bf T}$, there does not exist a coalition $S$ and a vector of types ${\bf T^{'S}}$ such that for all individuals $i$ in $S$,

\[ \rho_i({\bf T^{'S}}, {\bf T^S}) \geq  \rho_i({\bf T}) \]

\noindent and

\[ \rho_i({\bf T^{'S}}, {\bf T^S}) >  \rho_i({\bf T}). \]

\noindent for some $i \in S$.

\begin{proposition} \label{proptrianglegi} Let $g$ be a triangle. There does not exist a mechanism satisfying ex post group-incentive compatibility and efficiency.
\end{proposition}

%-------------------------------------------------------------

\subsection{Homophily}

In this last extension, we analyze the effect of homophily on friend-based ranking. Homophily is the tendency for individuals to form links with those who are similar to themselves. We show that moderate levels of homophily increase the likelihood of extracting a complete ranking whereas extreme homophily has the opposite effect.

In our setting, individuals differ only according to their private characteristic $\theta$ so we model homophily by dividing the community into two equal size groups according to $\theta$ (poor and rich, low ability and high ability, etc.). We assume individuals form within group links with probability $p_w$ and across group links with probability $p_a$ (which is \possessivecite{golub2012homophily} \textit{islands model} with two groups). Homophily increases as $p_w$ grows larger than $p_a$.

The matrix below shows the probability of links for a community of six individuals where individuals $\{ 1, 2, 3 \}$ are the low $\theta$ group and $\{ 4, 5, 6 \}$ are the high  $\theta$ group. The link between $1$ and $2$ is within group so this link forms with probability $p_w$ while the link between $1$ and $4$ is across group so forms with probability $p_a$.

\begin{singlespace}
\[ g = \bordermatrix{
~ & 1 & 2   & 3   & 4   & 5   & 6 \cr
1 & 0 & p_w & p_w & p_a & p_a & p_a \cr
2 &   & 0   & p_w & p_a & p_a & p_a \cr
3 &   &     & 0   & p_a & p_a & p_a \cr
4 &   &     &     & 0   & p_w & p_w \cr
5 &   &     &     &     & 0   & p_w \cr
6 &   &     &     &     &     & 0 \cr
}  \]
\end{singlespace}
\vspace{5mm}

The planner receives a comparison between a pair of individuals if there exists a third individual who is friends with both individuals in the pair. In the example above, the probability of observing the comparison $(1,2)$ is $\Pr[\exists \ k \neq 1,2: g_{1k}=g_{2k}=1] = 1-((1-p_w^2)(1- p_a^2)^3)$. We can generalize this calculation to a community of $n$ individuals with two equal size groups. For $i$ and $j$ within the same group, $\Pr[\exists \ k \neq i,j: g_{ik}=g_{jk}=1] = 1-(1-p_w^2)^{\frac{n}{2}-2}(1-p_a^2)^{\frac{n}{2}}$. For $i$ and $j$ in different groups, $\Pr[\exists \ k \neq i,j: g_{ik}=g_{jk}=1] = 1-(1-p_w p_a)^{n-2}$.

In a community of $n$ individuals, $n-1$ comparisons are necessary and sufficient to determine the complete ranking. The lowest is compared to the second lowest, the second to the third, and so on. Notice that all of these comparisons except for one are within group comparisons. The probability of observing these comparisons simultaneously defines the probability of deriving the complete ranking.
\begin{align*}
    \Pr[\text{Complete ranking}] = \big(1-(1-p_w^2)^{\frac{n}{2}-2}(1-p_a^2)^{\frac{n}{2}} \big)^{n-2} \big( 1-(1-p_w p_a)^{n-2} \big)
\end{align*}

Keeping $p_a$ constant, if we increase $p_w$, $\Pr[\text{Complete ranking}]$ will increase because we have raised the expected number of links.  We need to keep the expected number of links constant to isolate the impact of homophily. There are $n-2$ within group links for every $n$ across group links. Starting from a zero homophily base of $p=p_w=p_a$, we can analyze the impact of homophily by increasing $p_w$ and decreasing $p_a$ to keep the expected number of links constant. Let $p_w=p+\eta$, where $\eta$ is the homophily parameter. To keep the number of links constant, $p_a = p - \eta \frac{n}{n-2}$.

\begin{figure}[ht]
\centering
\includegraphics[width=0.8\textwidth]{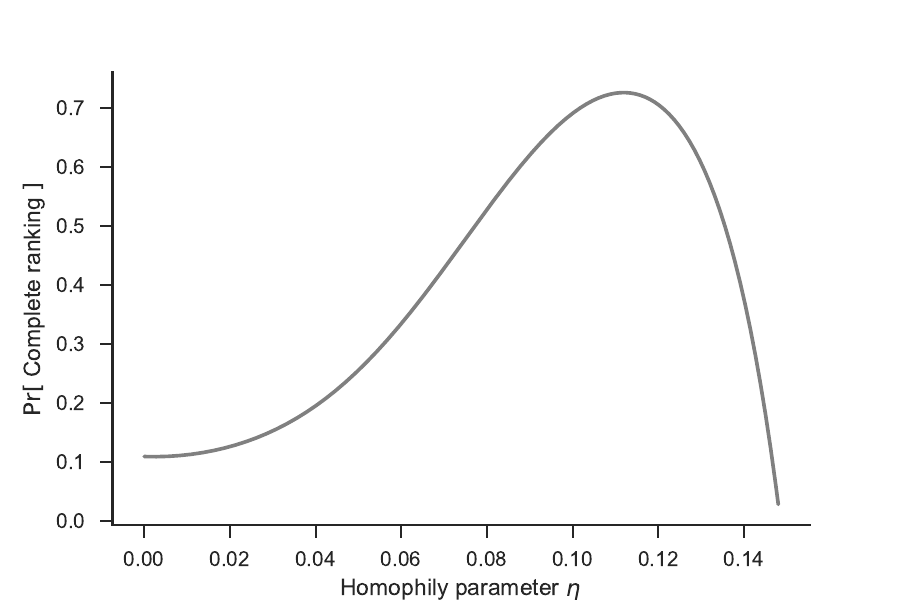}
\caption{Impact of homophily ($p=0.15$, $n=200$) }
\label{fig:homophily}
\end{figure}

From a base of $\eta = 0$ we can increase $\eta$ and observe how $\Pr[\text{Complete ranking}]$ responds. This is represented graphically in Figure \ref{fig:homophily} for a community size $n=200$ and a base probability of friendship $p=0.15$. Along the horizontal axis, as $\eta$ increases from 0 to around 0.1 the probability of observing the comparisons needed to derive the complete ranking increases. Intuitively, as homophily increases the probability that some individual $k$ is friends with two individuals in the same group increases while the probability that $k$ is friends with two individuals in different groups decreases. Since nearly all of the required comparisons are pairs within the same group, $\Pr[\text{Complete ranking}]$ rises with homophily.

However, notice that for $\eta>0.12$ the $\Pr[\text{Complete ranking}]$ drops sharply and approaches zero. The drop is driven by $p_a$ approaching zero. Even though the ranking relies primarily on within group links ($p_w$), there is always one comparison that depends on the across group links ($p_a$). The highest in the low group must be compared to the lowest in the high group. In our function above, as $p_a$ approaches zero the term $1-(1-p_w p_a)^{n-2}$ approaches zero and therefore $\Pr[\text{Complete ranking}]$ approaches zero.  Moderate levels of homophily improves friend-based ranking whereas extreme homophily worsens friend-based ranking.

%-------------------------------------------------------------
% Figure homophily
%-------------------------------------------------------------

%=============================================================

\section{Conclusion} \label{section:06}

This paper analyzes the design of mechanisms to rank individuals in communities in which individuals have only local, ordinal information on the characteristics of their friends. In these communities, pooling the information of all individuals may not be sufficient to obtain a complete ranking, and so we distinguish between completely informative communities and communities where incomplete rankings can be obtained.

In completely informative communities, we show that the planner can construct an ex post incentive-compatible and ex post efficient mechanism if and only if each pair of individuals is observed by a third individual, i.e., the individuals in each pair have a common friend. We use this insight to characterize the sparsest social network for which a complete ranking exists as constituting a ``friendship network" (or ``windmill network'') in the sense of  \citet{erdos1966problem}.

When the social network is not completely informative, we can generalize our theorem to any network. Even when the network may generate an incomplete ranking, the planner can construct an ex post incentive-compatible and ex post efficient mechanism if and only if each pair of friends has at least one friend in common.

Since the necessity arguments of our main theorems rely on self-comparisons, we analyze if the planner can do better when self-comparisons are discarded. We find two sufficient conditions. First, in bipartite networks, individuals on one side of the network can be used to rank individuals on the other side, resulting in an ex post efficient but incomplete ranking. Second, where all links are supported in triangles, the planner can use the congruence of comparisons to construct truthful rankings over any pair of individuals.

To the best of our knowledge, this is the first paper to analyze this intriguing theoretical problem---the design of a mechanism constructing a complete ranking when individuals have local, ordinal information based on a social network. In future work, we would like to further our understanding of the problem, by considering in more detail the difference between ordinal and cardinal information, between complete and coarse rankings, and between different concepts of implementation. On the empirical side, we believe that the rich data collected by \citet{alatas2016network} in Indonesia could be used to test the link between accuracy of community rankings and the architecture of the social network and the relation between individual rankings, aggregate rankings and the network structure. We plan to work on this full-fledged empirical study of the model  in the near future.

%=============================================================

\bibliographystyle{jpe}
\bibliography{references}

%=============================================================

% Start the appendix on a new page
\clearpage

% Begin the appendix here
\appendix

%-------------------------------------------------------------

\section{Proofs}

\noindent{\bf Proof of Lemma \ref{lemmacomp}}

The condition is obviously sufficient, as it guarantees that for any pair $(i,j)$ there exists an individual $k$ such that $t^k_{ij} \neq 0$, Hence the matrix generated by $(T^1,..,T^n)$ contains nonzero entries everywhere outside the diagonal. Conversely, suppose that there exists a pair of individuals $(i,j)$ who is observed by no other player and such that $g_{ij}=0$. Consider a realization of the characteristics such that $\theta_i$ and $\theta_j$ are consecutive in the ranking. No individual can directly compare $i$ and $j$. In addition, because there is no $k$ such that $\theta_k \in (\theta_i, \theta_j)$,  there is no $k$ such that $\theta_i \prec \theta_k \prec \theta_j$ or $\theta_j \prec \theta_k \prec \theta_i$. Hence the social network $g$ is not completely informative.

\bigskip

\noindent{\bf Proof of Theorem \ref{theocomp}}

\noindent {\em Sufficiency.} Suppose that for any pair of individuals $(i,j)$, there exists a third individual $k$ for whom $g_{ik}=g_{jk}=1$. We explicitly construct  the mechanism $\rho$. For any pair of individuals $i,j$, let  $r_{ij} \in \{0,1\}$ denote the bilateral ranking chosen by the planner.

First consider a pair of individuals $(i,j)$ who observe each other, $g_{ij}=1$. By assumption,  there are at least three reports on the ranking of $i$ and $j$. If all individuals transmit the same report $t_{ij}$, let $r_{ij}=t_{ij}$. If all individuals but one transmit the same report $t_{ij}$ and one individual reports $t^{'}_{ij} = - t_{ij}$, ignore the ranking $t^{'}_{ij}$ and let $r_{ij} = t_{ij}$. In all other cases, let $r_{ij}=1$ if and only if $i > j$.

Second consider a pair of individuals $(i,j)$ who do not observe each other, $g_{ij}=0$. By assumption, there exists at least one individual $k$ who observes them both. If there are at least three individuals who observe $i$ and $j$, use the same mechanism as above, $r_{ij}=t_{ij}$ if all individuals agree or one individual disagrees and $r_{ij}=1$ if and only if $i>j$ in all other cases.

If  two individuals observe $i,j$, select the individual with the highest index, $k$. Otherwise let $k$ be the unique individual who observes $i$ and $j$.  Consider the vector of announcements $\tilde{T}^{-k}$  where one disregards the announcements of individual $k$. Let $\succ_{\tilde{T}^{-k}}$ be the binary relation generated by letting $t_{ij}=1$ if and only if $t^l_{ij}=1$ for all $l \neq k$. Suppose that there exists a directed path of length greater or equal to $2$ between $i$ and $j$ in $\succ_{\tilde{T}^{-k}}$. If for all directed paths between $i$ and $j$ in $\succ_{\tilde{T}^{-k}}$, $i^0,..,i^L$ we have $t_{i^li^{l+1}} = 1$, then $r_{ij}=1$. If on the other hand for all directed paths between $i$ and $j$ in $\succ_{\tilde{T}^{-k}}$, $t_{i^li^{l+1}}=-1$, then $r_{ij}=-1$. In all other cases, let individual $k$  dictate the  comparison between $i$ and $j$, $r_{ij} = t^k_{ij}$.

As we have exhausted all possible situations, we have generated bilateral comparisons  $r_{ij}$ for every pair of individuals $i,j$. If they induce a transitive binary relation on $N$, let $\rho$ be the complete order generated by the  comparisons. Otherwise, consider all shortest cycles generated by the binary relation $r_{ij}$. If there exists a single individual $i$ who dictates at least two comparisons in all shortest cycles, individual $i$ is punished by setting  $\rho_i = 1$ and $\rho_j > \rho_k$ if and only if $j> k$ for all $j,k \neq i$.  If this is not the case, pick the arbitrary ranking where $\rho_i > \rho_j$ if and only if $i > j$.

We now show that the mechanism $\rho$ is ex post incentive-compatible and ex post efficient.

Suppose that all individuals except $k$ report their true type, and consider individual $k$'s incentive to report $T^{'k} \neq T^k$. We first note that there are circumstances where $k$'s report does not change the ranking chosen by the planner. This will happen on any comparison $(i,j)$ which is observed by at least three individuals, on any comparison $(i,j)$ where $k$ is not the individual with the highest index and on any comparison $(i,j)$ which can be constructed by using a path of length greater or equal to 2 using $\succ_{\tilde{T}^{-k}}$.  Hence we  focus attention on comparisons $(i,j)$ such that $k$ is the highest index individual who observes $i$ and $j$ and there is no directed path of length greater than or equal to 2 between $i$ and $j$ in $\succ_{\tilde{T}^{-k}}$.

We first show that individual $k$ cannot gain by making an announcement which generates an intransitivity in the planner's ranking $\rho$, as she will be detected as a deviator and be assigned to the lowest rank. To see this suppose that all
individuals $l \neq k$ announce the truth,  and that there exists a cycle in the ranking generated by the comparisons $r_{ij}$.  We first claim that the shortest cycles must be of order 3.

Suppose that there exists a cycle of order $L$, $i^0,i^1,..,i^L$. Because the community is completely informative, all binary comparisons are extracted by the planner, so that for any $l,m$, either $r_{i^li^m}=1$ or $r_{i^li^m}=-1$. Now consider three individuals $i^0,i^1,i^2$. If $r_{i^0i^2}=-1$, $i^0,i^1i^2i^0$ forms a cycle of order 3. If not, $r_{i^0i^2}=1$ and we  construct a cycle of order $L-1$, by eliminating the intermediary individual $i^1$. By repeating this argument, we either find cycles of order $3$ or end up reducing the initial cycle to a cycle of order 3.

Consider next a cycle of order 3 $ijli$ We claim that individual $k$ must dictate at least two comparisons in the cycle. If $k$ does not dictate any comparison, as all individuals but $k$ tell the truth, the ranking must be transitive. If $k$ dictates a single  comparison $(i,j)$ in the cycle but not the  comparisons $(j,l)$ and $(l,i)$, then there exists a directed path between $i$ and $j$ of length greater than or equal to 2  in $\succ_{T^{-k}}$, and individual $k$ cannot dictate the comparison between $i$ and $j$. We conclude that all shortest cycles are of order 3, and that in any cycle of length 3, individual $k$ must dictate at least two of the three  comparisons. The mechanism then identifies individual $k$ as the deviator, assigns rank $\rho(k)=1$ so that  $k$ cannot benefit from inducing a cycle.

Finally suppose that all comparisons $r_{ij}$ result in a transitive relation. We claim that the comparisons generated by $T^{-k}$  are sufficient to compute the rank of $k$. In fact, for any $i \neq k$, either $g_{ik}=1$ and there are at least three individuals making a report on the comparison, or $g_{ik}=0$ and the report on $(i,k)$ is  made by a common friend $l$. In both cases, the information contained in $T^{-k}$ is sufficient to construct the comparison $r_{ik}$. Hence $\rho_k$ is independent of the announcement $T^k$, concluding the proof that the mechanism is ex post incentive-compatible.

To show that the mechanism is ex post efficient notice that, when all individuals truthfully report their types, the rankings $r_{ij}$ induce a transitive relation, and yield the complete ranking generated by $\succ_{\bf T}$.

\medskip

\noindent{\em Necessity.}  Suppose that the social network $g$ satisfies the conditions of Lemma \ref{lemmacomp} but that there exists a pair of individuals $(i,j)$ who observe each other but are not observed by any third individual $k$. Consider a realization of the characteristics such that $\theta_i$ and $\theta_j$ are the two lowest characteristics. Let ${\bf T}_1$ be the type profile if $\theta_i < \theta_j$ and ${\bf T}_2$ the type profile if $\theta_j < \theta_i$.

By ex post efficiency, because the rankings generated by ${\bf T}_1$ and ${\bf T}_2$ are complete,

\begin{align*}
\rho_i({\bf T}_1) = \rho_j({\bf T}_2) = 1, & \\
\rho_i({\bf T}_2) = \rho_j({\bf T}_1) = 2. &
\end{align*}

Because there are only two announcements $t^i_{ij}$ and $t^j_{ij}$ on the link $(i,j)$, ex post incentive compatibility requires that individuals $i$ and $j$ cannot improve their ranking by changing their reports on the link $(i,j)$. Let $T_{-ij}$ denote the announcements on all links but link $ij$. We must have
\begin{align*}
\rho_i(T_{-ij}, t^i_{ij}=1, t^j_{ij}=-1) & = \rho_i({\bf T}_1) = 1, & \\
\rho_j(T_{-ij}, t^i_{ij}=1, t^j_{ij}=-1) & = \rho_j({\bf T}_2) = 1. &
\end{align*}

\noindent resulting in a contradiction as $i$ and $j$ cannot both be ranked at position 1.

\bigskip

\noindent{\bf Proof of Theorem \ref{theowindmill}}

 The proof of the Theorem differs from the well known proofs of the ``Friendship Theorem". We establish the Theorem through a sequence of claims. Let $\ell(g)$ be the number of links in the social network $g$.

\begin{claim} If the social network is completely informative, then every individual must have at least 2 friends.
\end{claim}

\noindent{\em Proof. } Let $d_i$ be the number of friends of individual $i$. As $g$ is connected, $d_i \geq 1$ for all $i \in N$. Suppose that $d_i=1$, and consider the unique friend $j$ of $i$. As $d_i=1$, there is no $k \neq j$ which is connected to $i$ and can draw a comparison between $i$ and $j$. Hence the network $g$ is not completely informative, establishing a contradiction.

\begin{claim} If for any $(i,j)$ there exists $k$ such that $g_{ik} g_{jk}=1$, then $\ell(g) \geq \frac{3(n-1)}{2}$ if $n$ is odd and $\ell(g) \geq \frac{3n}{2}-1$ if $n$ is even.
\end{claim}

\noindent{\em Proof.} Consider the following problem: For a fixed number of links $L$, compute the maximal number of comparisons of friends that can be generated by a social network $g$ when all nodes have degree contained in $[2, n-1]$. More precisely, let $(d_1,..,d_n)$ denote the degree sequence of $g$ with the understanding that $d_{i-1} \geq d_i$ for all $i=1,..,n$. Then consider the problem:
\begin{align*}
\max_{(d_1,...,d_n)} & \frac{d_1(d_1-1)}{2}+ \frac{d_2(d_2-1)}{2}+...+ \frac{d_n(d_n-1)}{2}& \\
 \text{ subject to } & 2 \leq d_i \leq n-1 \ \ \forall i \ , \\
& d_1 + d_2 +...+ d_n = 2L \ .
\end{align*}

Notice that the objective function $V(d_1,..,d_n) = \frac{d_1(d_1-1)}{2}+ \frac{d_2(d_2-1)}{2}+...+ \frac{d_n(d_n-1)}{2}$ is strictly increasing and convex in $(d_1,...,d_n)$.

Assume first that $n$ is odd. Then pick $L=\frac{3(n-1)}{2}$ and $d_1=n-1, d_2=..=d_n=2$. Because $V$ is strictly convex,
\begin{align*}
V(n-1,2,...,2)  &=  \frac{(n-1)(n-2)}{2} + n-1 \\
&= \frac{n(n-1)}{2} \\
&>  V(d_1,...,d_n)
\end{align*}
\noindent for any $(d_1,..,d_n) \neq (n-1,2,...2)$ such that $d_1+..+d_n = 3(n-1)$ and $d_i \geq 2$ for all $i$. Now $\frac{n(n-1)}{2}$ is the total number of  comparisons. So, as $V(d_1,..,d_n)$ is strictly increasing in $n$, the social network $g$ must contain at least $\frac{3(n-1)}{2}$ links for all comparisons to be constructed.

Assume next that $n$ is even. Pick $L=\frac{3n}{2}-1$ and $d_1=n-1,d_2=3,d_3=..=d_n=2$. Because $V$ is strictly convex,
\begin{align*}
V(n-1,3,2,...,2)  &=  \frac{(n-1)(n-2)}{2} + 3+n-2 \\
&= \frac{n(n-1)}{2}+1 \\
&>  V(d_1,...,d_n)
\end{align*}
\noindent for any $(d_1,..,d_n) \neq (n-1,3,2,...2)$ such that $d_1+..+d_n = 3n-2$ and $d_i \geq 2$ for all $i$.

In addition notice that for $L'= \frac{3n}{2}-2$,
\begin{align*}
V(n-2,2,2,...,2)  & =  \frac{(n-2)(n-3)}{2} + n-1 \\
& = \frac{(n-2)^2-n}{2} \\
& > V(d_1,...,d_n)
\end{align*}
\noindent for any $(d_1,..,d_n) \neq (n-2,2,2,...2)$ such that $d_1+..+d_n = 3n-4$ and $d_i \geq 2$ for all $i$.

Hence, the maximum of $V_i$ is smaller than $\frac{n(n-1)}{2}$ when $\ell(g)= \frac{3n}{2}-2$ and greater than $\frac{n(n-1)}{2}$ when $\ell(g)= \frac{3n}{2}-1$, establishing that the social network $g$ must contain at least $\frac{3n}{2}-1$ links for all comparisons to be constructed.

Next we observe that the friendship network and the modified windmill network generate all comparisons.

\begin{claim} If $n$ is odd, the friendship network containing exactly $\frac{3(n-1)}{2}$ links,  generates all comparisons. If $n$ is even, the windmill with sails of size $2$ and one sail of size $3$ with an additional link, containing exactly $\frac{3n}{2}-1$ links, generates all comparisons.
\end{claim}

\noindent{\em Proof.} The hub of the network, node $n_h$, provides the  comparisons between all other $(n-1)$ nodes. If $n$ is odd, in any petal $(i,j)$, $i$ provides the  comparison between $j$ and $n_h$ and $j$ provides the  comparison between $i$ and $n_h$. If $n$ is even, in any sail of size $2$, $(i,j)$,$i$ provides the  comparison between $j$ and $n_h$ and $j$ provides the  comparison between $i$ and $n_h$. In the unique sail of size 3, $(i,j,k)$, $i$ provides the  comparison between $j$ and $n_h$, $j$  provides the  comparisons between $i$ and $n_h$ and $k$ and $n_h$ and $k$ provides a (redundant)  comparison between $j$ and $n_h$.

Finally we establish that the friendship network and the modified windmill network are the only network architectures generating all comparisons with the minimal number of edges.

\begin{claim} If $n$ is odd, the friendship network is the only network with degree sequence $(n-1,2,...,2)$. If $n$ is even, the modified windmill network with $\frac{n}{2}-2$ sails of size $2$ and one sail of size $3$ with an additional link is the only network with degree sequence $(n-1,3,2,..2)$.
\end{claim}

\noindent{\em  Proof.} Let $n$ be odd. Because one node has degree $n-1$, the network is connected and this node is a hub. All other nodes must be connected to the hub, and if they have degree $2$, they must be mutually connected to one other node. Let $n$ be even. The same argument shows that all nodes with degree $2$ must be connected to the hub and one other node. These nodes are mutually connected except for the petal of size 3, where one node is connected to the two other nodes in the sail.

\bigskip

\noindent{\bf Proof of Theorem \ref{theocomp2}}

\noindent{\em Sufficiency.}  The mechanism $\rho$ is constructed exactly in the same way as the mechanism in Theorem \ref{theocomp}.
We now prove that the mechanism satisfies ex post efficiency and incentive compatibility.

Consider any pair $(i,j)$ such that $i \succ_{\bf T} j$. There must exist a sequence of  comparisons $(i,i^1,..,i^t,..,i^T,j)$ such that $h_{i^{t-1}i^t} =1$ and $i^{t-1} \succ_{\bf T} i^t$. For any of these pairs, we must have $r_{i^{t-1}i^t} = 1$ and hence, because the announcement ${\bf T}$ generates a transitive partial order, $\rho(i^{t-1}) > \rho(i^t)$. But this implies that $\rho(i) > \rho(j)$, establishing that the mechanism satisfies ex post efficiency.

Next we show that the mechanism is ex post incentive-compatible. Consider individual $k$'s incentive to change his announcement on a link $ij$ when all other individuals tell the truth. If the link $ij$ is supported, this change does not affect the outcome of the mechanism. So consider an unsupported link $ij$ and let individual $k$ be the highest index individual observing $i$ and $j$. Suppose that all individuals $l \neq k$ announce the truth, so that $\succ_{\tilde{T}^{-k}} = \succ_{T^{-k}}$. We first show that individual $k$ cannot gain by making an announcement which generates cycles in the ranking $r_{ij}$.

Suppose that the binary relation generated by $r_{ij}$ exhibits a cycle $i^0i^1...i^L$

By the same argument as in the proof of Theorem \ref{theocomp}, individual $k$ must dictate at least two comparisons in the cycle. We will show that the initial cycle must contain a cycle of order 3. Suppose that the initial cycle has order greater than or equal to $4$. Let $ij$ and $lm$ be two  comparisons dictated by individual $k$. Suppose first that $j \neq l$. As individual $k$ observes both $(i,j)$ and $(l,m)$, he also observes both $i$ and $l$. Hence $i$ and $l$ must be compared under $r$ and  either $r_{il}=1$ or $r_{il}=-1$. Now if $r_{il}=-1$, one can construct a shorter cycle by replacing the path $lm..i$ by the path $li$. If $r_{il}=+1$, one can construct a shorter cycle by replacing the path $ij,,l$ by the path $il$. Next suppose that $j=l$ so that the two comparisons $(i,j)$ and $(l,m)$ are adjacent in the cycle. Again because individual $k$ observes both $i$ and $m$, then $i$ and $m$ must be compared under $r$ and either $r_{im}=+1$ or $r_{im}=-1$. If $r_{im}=+1$, one can construct a shorter cycle by replacing $ijm$ with $im$. If $r_{im}=-1$, one can construct a cycle of order 3 $ijmi$.

We conclude that if the binary relation $r$ exhibits a cycle,  all shortest cycles are of order 3 and individual $k$ dictates at least two of the comparisons in all cycles. Hence, individual $k$ can be detected as a deviator and has no incentive to make an announcement generating a cycle in $r$,  as he will be punished and assigned the lowest rank.

We finally assume that the ranking generated by $r$ is acyclic and show that the rank of individual $k$ must remain the same if he changes his report on any pair $(i,j)$. Notice first that, if $k$ dictates the ranking between $i$ and $j$, either individuals $ i$ and $j$ are either both ranked above or below $k$ by the reports $T^{-k}$. If this were not the case, there would exist a path of length $2$ between $i$ and $j$ in $T^{-k}$, contradicting the fact that $k$ dictates the ranking on that comparison.

Now let $J = (i_1,j_1),...,(i_l,j_l),..(i_L,j_L)$ be the pairs on which $k$ is a dictator and let $J^+$ denote the set of pairs $(i_l,j_l)$ such that $k \succ_{T^{-k}}i_l,j_l$ and $J^{-}$ the set of pairs such that $k \prec_{T^{-k}} i_l,j_l$. Let ${\bf T'} = (T^{-k}, T^{'k})$ the announcement obtained when $i$ changes his report on some of the pairs in $J$ while keeping a transitive partial order. For any $m$ such that $k \succ_{T{-k}} m$, $k \succ_{\bf T} m$ and $k \succ_{\bf T'} m$. Hence $\rho_k({\bf T}) > \rho_m({\bf T})$ and $\rho_k({\bf T'}) > \rho_m({\bf T'})$. Similarly, for any $m$ such that $k \prec_{T{-k}} m$, $k \prec_{\bf T} m$ and $k \prec_{\bf T'} m$. Hence $\rho_k({\bf T}) < \rho_m({\bf T})$ and $\rho_k({\bf T'}) < \rho_m({\bf T'})$. We also have, for any $m \in J^+$, $\rho_k({\bf T}) > \rho_m({\bf T})$ and $\rho_k({\bf T'}) > \rho_m({\bf T'})$. For any $m \in J^-$, $\rho_k({\bf T}) < \rho_m({\bf T})$ and $\rho_k({\bf T'}) < \rho_m({\bf T'})$. Next consider $m$ such that $k \bowtie_{T^{-k}} m$ and $m \notin J$. If $m \prec_{T^{-k}}i_l$ for some $i_l \in J^+$, then $k \succ_{\bf T} m$ and $k \succ_{\bf T'} m$ so that $\rho_k({\bf T}) > \rho_m({\bf T})$ and $\rho_k({\bf T'}) > \rho_m({\bf T'})$. Similarly, if $m \succ_{T^{-k}} i_l$ for some $i_l \in J^-$, then $k \prec_{\bf T} m$ and $k \prec_{\bf T'} m$ so that $\rho_k({\bf T}) < \rho_m({\bf T})$ and $\rho_k({\bf T'}) < \rho_m({\bf T'})$. Finally, if $m \succ_{T^{-i}} i_l \forall i_l \in J^+, m \prec_{T^{-k}} i_l \forall i_l \in J^-$ and $k \bowtie_{T^{-k}} m$, then $k \bowtie_{\bf T} m$ and $k \bowtie_{\bf T'} m$. Whenever $k \bowtie_{\bf T} m$ and $k \bowtie_{\bf T'} m$, then the ranking between $k$ and $m$ is independent of the type profile. Hence, in all cases the ranking between $k$ and $m$ is identical under ${\bf T}$ and ${\bf T'}$. This argument completes the proof that the mechanism satisfies ex post incentive compatibility.

\medskip

\noindent {\em Necessity.} Let $i$ and $j$ be two nodes such that $g_{ij}=1$ but there is no $k$ such that $g_{ik} = g_{jk} =1$. Consider a realization of the characteristics such that $\theta_i$ and $\theta_j$ are the two lowest characteristics. Partition the set of individuals different from $i$ and $j$ according to their distance to the node $i$. Let $A_i^1 = \{a| a \neq j, d(a,i) =1\}, A_i^2 = \{a|  d(a,i) =2\}$, $A_i^m = \{a | d(a,i) = m\}$. Fix the characteristics of the individuals in such a way that $\theta_k > \theta_l$ if $d(k,i) > d(l,i)$.

Consider two type profiles ${\bf T}_1$ and ${\bf T}_2$ which agree on all comparisons except that $\theta_i < \theta_j$ in ${\bf T}_1$ and $\theta_j < \theta_i$ in ${\bf T}_2$. Clearly, for any $k \neq i,j$, if $k \succ_{{\bf T}_1} j$ then $k \succ_{{\bf T}_1} i$, as $i$ and $j$ can be compared under ${\bf T}_1$. Similarly, if $k \succ_{{\bf T}_2} i$ then $k \succ_{{\bf T}_2} j $. Furthermore, as $\theta_i$ and $\theta_j$ are the two smallest characteristics, all individuals $k$ which can be compared to $i$ and $j$ have higher rank than $i$ and $j$.

We now claim that all individuals can be compared to $i$. Pick an individual $k$ at a distance $d(k,i)$ from $i$. Consider a shortest path $k=k^0,...i=k^m$ from $k$ to $i$. Then  $k^{m-1} \in A_i^1, k^{m-2} \in A_i^2...k^0 \in A_i^{d(k,i)}$. Hence $\theta_{k^1} < \theta_{k^2}<... \theta_{k^m}= \theta_k$. As $g_{k^1k^2} = g_{k^2k^3} =..=g_{k^{m-1} k^m}=1$, we must have $t_{k^1k^2} = t_{k^2k^3} =..=t_{k^{m-1} k^m}=-1$, so that $k$ can be compared to $i$ and has a higher rank than $i$ and $j$.

Now, by ex post efficiency,

 \begin{align*}
\rho_i({\bf T}_1) = \rho_j({\bf T}_2) = 1, & \\
\rho_i({\bf T}_2) = \rho_j({\bf T}_1) = 2. &
\end{align*}

Because there are only two announcements $t^i_{ij}$ and $t^j_{ij}$ on the link $(i,j)$, ex post incentive compatibility requires that individuals $i$ and $j$ cannot improve their ranking by changing their reports on the link $(i,j)$. Let $T_{-ij}$ denote the announcements on all links but link $ij$. We must have
\begin{align*}
\rho_i(T_{-ij}, t^i_{ij}=1, t^j_{ij}=-1) & = \rho_i({\bf T}_1) = 1, & \\
\rho_j(T_{-ij}, t^i_{ij}=1, t^j_{ij}=-1) & = \rho_j({\bf T}_2) = 1, &
\end{align*}

\noindent resulting in a contradiction as $i$ and $j$ cannot both be ranked at position 1.

\bigskip

\noindent{\bf Proof of Proposition \ref{proconn}}

 We first prove the following Claim.
\begin{claim} The comparison network is connected if and only if for all $i,j \in N$, there exists an even walk between $i$ and $j$.
\end{claim}
\bigskip

\noindent{\em Proof.}  Suppose first that $h$ is connected. Pick any two nodes $i,j \in N$ and a walk $i=i^0,...,i^m=j$ in $h$. By definition, for any $(i^k,i^{k+1})$ in the walk, there exists $j^k \in N$ such that $i^k, i^{k+1} \in N_{j^k}$. But this implies that there exists a walk in $g$ connecting $i$ to $j$ given by $i^0, j^0, i^1, j^1,...,i^{m-1}, j^{m-1}, i^m$.\footnote{Note that this walk is not necessarily a path even if the initial walk in $h$ is a path, as the same node $j^k$ can be used several times in the walk.} This walk contains an even number of edges, proving necessity of the claim.

\bigskip

Next suppose that $h$ is not connected and let $i$ and $j$ be two nodes in different components of $h$. We want to show that all walks between $i$ and $j$ in $g$  are odd. Consider first a path between $i$ and $j$. If the path is even, there exists a sequence of nodes $i=i^0, i^1,..,i^m=j$ where $m=2l$ is even such that $g_{i^k, i^{k+1}}=1$ for all $k$. But then, for any $l=0, \frac{m}{2}-1$, $h_{i^{2l}, i^{2l+1}} = 1$, and hence there exists a path $i=i^0,i^2,...,i^m=j \in f$, contradicting the fact that $i$ and $j$ belong to two different components in $h$. Hence all paths between $i$ and $j$ are odd. If there exists an even walk between $i$ and $j$ in $g$, it must thus involve an odd cycle starting at $i$ or starting at $j$. Without loss of generality, suppose that there exists an odd cycle starting at $i$, $i=i^0,...,i^m=i$, where $m=2l+1$ is odd. Consider any even path between $i$ and $j$, where we index $i=i^m,...i^r=j$ and $r=2p$ is even. We construct a path in $h$ between $i$ and $j$ as follows. Because $m$ is odd, we first construct the sequence of connected nodes in $h$,  $i=i^{2l+1}, i^{2l-1}, i^{2l-3},..,i^1$. Because $i$ is connected in $g$ both to $i^1$ and $i^{m+1}$, we then link $i^1$ to $i^{m+1}$ in $h$. Now $m+1$ is even, so we can use the path between $i$ and $j$ to construct a sequence $i^{m+1},..,i^r=j$ in $h$. Concatenating the two sequences, we construct a sequence $i, i^{m-2},...,i^1, i^{m+1},..,i^r=j$ in $h$, contradicting the fact that $i$ and $j$ belong to two different components in $h$. Hence if $h$ is not connected, there exists a pair of nodes $i,j$ such that all walks between $i$ and $j$ are odd, proving the necessity of the claim.

We now prove the second claim

\begin{claim} For all $i,j \in N$ there exists an even walk between $i$ and $j$ if and only if $g$ is not bipartite.
\end{claim}

\noindent{\em Proof.} Suppose that $g$ is bipartite with sets $A$ and $B$. As $N \geq 3$, at least one of the two sets has more than one element. Pick $i,j$ such that $i \in A$ and $j \in B$, then we claim that all walks between $i$ and $j$ must be odd. Any walk between $i$ and $j$ must contain an even number of edges alternating between nodes in $A$ and $B$ and a single edge between a node in $A$ and a node in $B$. Hence the total number of edges must be odd, proving the necessity of the claim.

\bigskip
Conversely, suppose that there exists a pair of nodes $i,j$ such that all walks between $i$ and $j$ are odd. Consider the sets of nodes $A=\{k| \delta(i,k) \mbox{ is even} \}$ and $B = \{k| \delta (i,k) \mbox{  is odd } \}$, where $\delta(i,k)$ denotes the geodesic distance between $i$ and $k$ in the graph. We first claim that if $k \in A$, {\em all walks between $i$ and $k$ must be even}. Suppose not, then there exist two different walks between $i$ and $k$, one $w_1$ which is even (the shortest path between $i$ and $k$) and one $w_2$ which is odd. Pick one particular path $p$ between $k$ and $j$. If this path is odd, then the walk between $i$ and $j$ containing $w_2$ followed by $p$ is even, contradicting the assumption. If the path is even, then the walk between $i$ and $j$ containing $w_1$ followed by $p$ is even, contradicting the assumption again. Hence all walks between $i$ and nodes in $A$ are even and all walks between $i$ and nodes in $B$ are odd. Next notice that there cannot be any edge between nodes in $A$. Suppose by contradiction that there exists an edge between $k$ and $l$ in $A$, and consider a walk between $i$ and $k$, $w_1$ followed by the edge $kl$. This forms an odd walk between $i$ and $l$, contradicting the fact that all walks between $i$ and $l$ must be even. Hence, there is no edge between nodes in $A$ and similarly no edge between nodes in $B$, showing that the graph $g$ is bipartite.

\bigskip

\noindent{\bf Proof of Proposition \ref{probipartite}}

 Consider a mechanism where all individuals in $A$ are ranked above individuals in $B$.  For any two individuals $i$ and $j$ in $A$, let $\rho(i) > \rho(j)$ if $i \succ_{\bf T} j$. If $i \bowtie_{\bf T} j$ or if the reports on $i$ and $j$ are incompatible, construct an arbitrary ranking by letting $\rho(i) > \rho(j)$ if and only if $i > j$. Similarly, for any two individuals $i$ and $j$ in $B$, let $\rho(i) > \rho(j)$ if $i \succ_{\bf T} j$. If $i \bowtie_{\bf T} j$ or if the reports on $i$ and $j$ are incompatible, let $\rho(i) > \rho(j)$ if and only if $i > j$.

We will show that the mechanism satisfies ex post incentive compatibility and efficiency.

The mechanism satisfies strategy-proofness, a stronger incentive compatibility notion than ex post incentive compatibility. Consider an individual $i$ in $A$. Then we claim that if $t^i_{jk} \neq 0$ it must be that both $j$ and $k$ are in $B$. To see this notice that as $g$ is bipartite it does not contain any triangle. Hence no self-comparison can be supported by a third individual, and hence $t^i_{ij} =0$ for all $j \neq i$. The only case where $t^i_{jk} \neq 0$ is thus when $g_{ij} g_{ik} = 1$ and $j, k \in B$. Hence, by changing his report $t^i_{jk}$, individual $i$ can only affect the ranking of individuals in $B$. As all individuals in $B$ are ranked below individuals in $A$, this does not affect the rank of individual $i$, and hence individual $i$'s ranking is independent of his announcement, proving that the mechanism is strategy-proof.

Finally, notice that by construction, the mechanism $\rho$ achieves an ex post efficient ranking separately on each of the two components $A$ and $B$. by Proposition \ref{proconn}, the comparison network $h$ is disconnected into two components $A$ and $B$. Hence the mechanism $\rho$ is also ex post efficient.

\bigskip

\noindent{\bf Proof of Proposition \ref{proptriangleds}}

We first establish the following simple general claim:

\begin{claim} \label{claimsp}If $\rho$ is strategy-proof, $\rho_i(T^i, \hat{T}^{-i})= \rho^i(T^{'i}, \hat{T}^{-i})$ for all $i, T^i, T^{'i}, \hat{T}^{-i}$.
\end{claim}

\noindent{\em Proof}. Suppose by contradiction that there exists $i, T^i, T^{'i}, \hat{T}^{-i}$ such that $\rho_i(T^i, \hat{T}^{-i})> \mu^i(T^{'i}, \hat{T}^{-i})$. Let $T^{'i}$ be the true type of individual $i$. Then,  individual $i$ has an incentive to announce $T^i$, contradicting the fact that $\rho$ is strategy-proof.

Consider next two vectors of  types:

\begin{itemize}
    \item $\mathbf{T}_1$: $t_{ij}=t_{jk}=t_{ik}=1$
    \item $\mathbf{T}_2$: $t_{ij}=-1, t_{jk}=1, t_{ik}=-1$
\end{itemize}

As the mechanism is ex post efficient, it must assign ranks $\rho_i(\mathbf{T}_1)=3, \rho_j(\mathbf{T}_1)=2, \rho_k(\mathbf{T}_1)=1, \rho_i(\mathbf{T}_2)=1, \rho_j(\mathbf{T}_2)=3, \rho_k(\mathbf{T}_2)=2.$

Now let $t^i$ denote the announcement $t^i_{ij}=t^i_{jk}=t^i_{ik}=1$ and $t^{'i}$ the announcement $t^i_{ij}=-1, t^i_{jk}=1, t^i_{ik}=-1$. By Claim \ref{claimsp},
\begin{align*}
\rho_i(t^{'i}, t^j,t^k) & =  \rho_i(t^i,t^j,t^k) & = 3, \\
\rho_j(t^i, t^{'j}, t^k) & = \rho_j(t^i,t^j,t^k) & = 2. \\
\rho_k(t^{'i}, t^{'j}, t^k) & = \rho_k(t^{'i}, t^{'j}, t^{'k}) & = 2.
\end{align*}

Hence we conclude that, at $(t^{'i}, t^{'j}, t^k)$ either $\rho_i=3, \rho_j=1$ or $\rho_i=1, \rho_j=3$. But $\rho_j=3$ is impossible, as, by claim \ref{claimsp},  $\rho_j(t^{'i}, t^{'j}, t^k) = \rho_j(t^{'i}, t^j,t^k)$ and $\rho_j(t^{'i}, t^j,t^k) \neq \rho_i(t^{'i}, t^j, t^k) = 3$. Hence we conclude that

\begin{equation} \rho_i(t^{'i}, t^{'j},t^k) = 3, \rho_j(t^{'i}, t^{'j},t^k) = 1, \rho_k(t^{'i}, t^{'j},t^k) = 2.
    \label{eq1}
\end{equation}

A similar reasoning shows that

\[ \rho_j (t^{'i}, t^j, t^{'k}) = 3, \]

\noindent and hence either $\rho_i = 2, \rho_k = 1$ or $\rho_i = 1, \rho_k=3$ at $(t^{'i}, t^j, t^{'k})$. But $\rho_i(t^{'i}, t^j, t^{'k}) = \rho_i(t^i, t^j, t^{'k}) \neq \rho_k(t^i,t^j,t^{'k}) = \rho_k(t^i,t^j,t^k) = 1$. So we conclude that

\
\begin{equation} \rho_i(t^{'i}, t^j,t^{'k}) = 2, \rho_j(t^{'i}, t^j,t^{'k}) = 3, \rho_k(t^{'i}, t^j,t^{'k}) = 1. \label{eq2}
\end{equation}

Now,   $\rho_j(t^{'i}, t^j,t^k) = \rho_j(t^{'i}, t^{'j},t^k)$.  By equation \ref{eq1},

\[\rho_j(t^{'i}, t^{'j},t^k) = 1\],

\noindent so that $\rho_j(t^{'i}, t^j,t^k) = 1$. As $\rho_i(t^{'i},t^j,t^k) = \rho_i(t^i,t^j,t^k)=1,$

\[ \rho_k(t^{'i}, t^j, t^k) = 2.\]

\noindent Similarly, $\rho_k(t^{'i}, t^j, t^k) = \rho_k(t^{'i}, t^j, t^{'k})$ and by equation \ref{eq2},

\[\rho_k(t^{'i}, t^j, t^{'k}) =1\]

\noindent so that

\[ \rho_k(t^{'i}, t^j, t^k) = 1.\]

\noindent establishing a contradiction.

\bigskip

\noindent{\bf Proof of Proposition \ref{coarse}}

 If individuals strictly prefer being ranked at $\rho(i)=2$ to being ranked at $\rho(i)$, the necessity part of the proof of Theorem \ref{theocomp} shows that whenever there exists a pair of individuals who are not observed by a third individual, there cannot exist an ex post incentive-compatible and efficient mechanism.

Conversely, if individuals are indifferent between being ranked at $\rho(i)=1$ and $\rho(i)=2$, let $\rho(i)=1$ and $\rho(j) =2$ whenever $i$ and $j$ are the only two individuals observing the ranking between $i$ and $j$ and $t^i_{ij} \neq t^j _{ij}$. This guarantees that individuals have no incentive to send conflicting reports, and hence that this mechanism, completed by the mechanism constructed in the sufficiency part of Theorem \ref{theocomp}, satisfies ex post incentive compatibility and efficiency.

\bigskip

\noindent{\bf Proof of Proposition \ref{proptrianglegi}}

Consider a vector of announcements where all three individuals agree on $t_{13}=-1, t_{23}=-1, t_{12}=1$. By ex post efficiency, $\rho(1)=2, \rho(2)=1, \rho(3)=3$. We claim that ex post group-incentive compatibility implies that, whenever individual $3$ announces $t^3_{13}=-1, t^3_{23}=-1, t^3_{12}=1$, the rank of individual $1$ must be different from $3$. If that were not the case, there would exist an announcement $(t^{'1}, t^{'2})$ for individuals $1$ and $2$ resulting in a rank $\rho(1) = 3 > 2, \rho(2) \geq 1$, contradicting ex post group-incentive compatibility. By a similar reasoning, whenever individual $1$ announces $t^1_{12} =1, t^1_{13}=1, t^1_{23} =1$, the rank of individual $2$ must be different from $3$. Finally, when individual $2$ announces $t^2_{12}=-1, t^2_{23}=1, t^2_{13}=-1$, the rank of individual $3$ must be different from $3$.

So consider the announcement $t^1=(t^1_{12} =1, t^1_{13}=1, t^1_{23} =1),t^2=(t^2_{12}=-1, t^2_{23}=1, t^2_{13}=-1),t^3=(t^3_{13}=-1, t^3_{23}=-1, t^3_{12}=1)$. For this announcement, neither of the three individuals can be in position $3$, completing  the proof of the Proposition.

\end{document}